\documentclass{aa}
\usepackage[varg]{txfonts}
\usepackage{placeins}
\usepackage{pdflscape}

\usepackage{natbib}
\bibpunct{(}{)}{;}{a}{}{,}

\defcitealias{2009A&AMordasinia}{M09a}

\defcitealias{2009A&AMordasinib}{M09b}

\defcitealias{2011MayorArxiv}{M11}

\defcitealias{2021AAEmsenhuberA}{NGPPS I}
\def\paperone{\citetalias{2021AAEmsenhuberA}}

\defcitealias{2021AAEmsenhuberB}{NGPPS II}
\def\papertwo{\citetalias{2021AAEmsenhuberB}}

\usepackage{longtable}

\usepackage{babel}

\usepackage{amssymb, amsmath}
\usepackage{wasysym}

\usepackage{siunitx}
\DeclareSIUnit\erg{erg}
\DeclareSIUnit\year{yr}
\DeclareSIUnit\au{au}

\def\mearth{M_\oplus}
\def\msun{M_\odot}
\def\mj{M_{\textrm{\tiny \jupiter}}}

\def\mstar{M_\star}
\def\mplanet{M_\mathrm{planet}}

\def\rplan{R_\mathrm{plan}}

\def\nobs{N_{\rm obs}}

\usepackage{hyperref}
\hypersetup{pdfauthor=A. Emsenhuber et al., pdftitle=The New Generation Planetary Population Synthesis (NGPPS). VII., colorlinks=true, urlcolor=blue, linkcolor=blue, citecolor=blue}

\title{The New Generation Planetary Population Synthesis (NGPPS)}
\subtitle{VII. Statistical comparison with the HARPS/Coralie survey}
\titlerunning{The New Generation Planetary Population Synthesis (NGPPS). VII.}

\author{Alexandre Emsenhuber\inst{\ref{unibe},\ref{usm},\ref{ualpl}} \and Christoph Mordasini\inst{\ref{unibe}} \and Michel Mayor\inst{\ref{unige}} \and Maxime Marmier\inst{\ref{unige}} \and St\'ephane Udry\inst{\ref{unige}} \and Remo Burn\inst{\ref{mpia}} \and Martin Schlecker\inst{\ref{uaas}} \and Lokesh Mishra\inst{\ref{ibm}} \and Yann Alibert\inst{\ref{unibe}} \and Willy Benz\inst{\ref{unibe}} \and Erik Asphaug\inst{\ref{ualpl}}}
\authorrunning{A.~Emsenhuber et al.}

\institute{
Weltraumforschung und Planetologie, Universität Bern, Gesellschaftsstrasse 6, 3012 Bern, Switzerland\\ \email{alexandre.emsenhuber@unibe.ch}\label{unibe}
\and
	Universitäts-Sternwarte München, Ludwig-Maximilians-Universität München, Scheinerstraße 1, 81679 München, Germany
	\label{usm}
	\and
	Lunar and Planetary Laboratory, University of Arizona, 1629 E. University Blvd., Tucson, AZ 85721, USA\label{ualpl}
	\and
	Observatoire Astronomique de l'Université de Genève, 51 ch. de Pegasi, 1290 Versoix, Switzerland
	\label{unige}
	\and
	Max Planck Institute for Astronomy, Königstuhl 17, 69117 Heidelberg, Germany\label{mpia}
	\and
 	Steward Observatory, University of Arizona, 933 N. Cherry Ave., Tucson, AZ 85721, USA\label{uaas}
    \and
    IBM Research, 8803 Rüschlikon, Switzerland\label{ibm}
}

\date{Received 4 October 2024 / Accepted 27 May 2025}

\abstract
{Planetary population synthesis is a tool that is used to better understand the key processes of planet formation at the statistical level.}
{We seek to quantify the fidelity with which modern population syntheses reproduce observations in view of their use as predictive tools.}
{We compared synthetic populations from the Generation 3 Bern Model of Planet Formation and Evolution (core accretion, solar-type host stars) and the HARPS/Coralie radial velocity sample. We biased the synthetic planet population according to the completeness of the observed data. We then performed quantitative statistical comparisons and systematically identified agreements and  differences.}
{Our nominal population reproduces many of the main features of the HARPS planets, such as two main groups of planets in the mass-distance diagram (close-in sub-Neptunes and distant giants), a bimodal mass function with a less populated `desert', an observed mean multiplicity of about 1.6, and several key correlations regarding the stellar metallicity dependency, the period ratio distribution, and the eccentricity distribution. Considering that the model was not optimised beforehand to reproduce any particular survey, this indicates that some of the important physical processes governing planetary formation could be captured. The remaining discrepancies that can be quantified thanks to the population synthesis approach point to areas that are not fully captured in the model. For instance, we find that the synthetic population has 1) in absolute terms too many planets by \SI{\approx70}{\%}, 2) a `desert' that is too deep by \SI{\approx60}{\%}, 3) a relative excess of giant planets by \SI{\approx40}{\%}, 4) planet eccentricities that are on average too low by a factor of about two (median of \num{0.07} versus \num{0.15}), and 5) a metallicity effect that is too weak. Finally, the synthetic planets are overall too close to the star compared to the HARPS sample.
The differences allowed us to find model parameters that better reproduce the observed planet masses, for which we computed additional synthetic populations.
We find that decreasing the planet formation efficiency by increasing the planetesimal size re-balances the number of sub-Neptunes versus giant planets. Changing the efficiency of gas-driven migration also affects the sub-Neptune to giant planet ratio, with lower migration rates resulting in more giant planets and fewer sub-Neptunes.}
{However, only modifying the model parameters seems to be insufficient for the model to fully reproduce both the observed mass and distance distributions at the same time. Instead, physical processes appear to be missing. Planets may originate on wider orbits than our model predicts. Mechanisms leading to higher eccentricities and slower disc-limited gas accretion also seem necessary. We also advocate that theoretical models should make a quantitative, rather than merely a qualitative, comparison between the many current and future large surveys and theoretical results to better understand the origins of planetary systems.}


\begin{document}

\maketitle

\section{Introduction}

The formation and evolution of planetary systems is far from being understood. On the observational side, data exist for early processes such as dust growth and radial drift \citep[e.g.][based on ALMA data]{2020AATrapmanA,2023PPVIIManara}. However, direct observation of forming planets is difficult. Only a handful of cases at most are known, such as the PDS 70 system, with PDS 70b \citep{2018AAKeppler,2018AAMueller} and PDS 70c \citep{2019NatAsHaffert}, and AB Aurigae b \citep{2022NatAsCurrie}. Other recent studies have to date been unable to find additional accreting planets \citep{2019AACugno,2020AAZurlo}.

On the theoretical side, many mechanisms that are supposed to take place during the later stages of planetary formation have been proposed. This includes how solids grow from pebbles and planetesimals to planets \citep{1992IcarusGreenzweigLissauer,1993IcarusIdaMakino,2002IcarusOhtsuki,2003IcarusThommes,2010AAOrmelKlahr,2018NatAsAlibert}; how giant planets accrete their envelope \citep{1996IcarusPollack,2000IcarusBodenheimer,2012MNRASAyliffeBate,2014AAMordasiniA,2020ApJVallettaHelled,2022AASchib,2023AANelson}; and how planetary migration affects the architecture of the systems \citep{2016SSRvBaruteau,2022AAChrenko,2023PPVIIPaardekooper,2023PPVIIWeiss}. However, the lack of direct observations of these processes makes it difficult to constrain them.

An alternative approach is to compare the end product from theoretical models (namely the planetary systems) with observations. However, it is difficult to reproduce single exoplanetary systems. While the models can predict many observables, there are usually only a few quantities that have actually been observed (the orbits, masses, or radii of each planet), and even less is known about the initial conditions of a specific system (e.g. the disc metallicity, which is likely related to the stellar value, \citealp{2016ApJGaspar}).

Instead, one can leverage the thousands of discovered planets \citep{2021ARAAZhuDong,2023PPVIIWeiss} and protoplanetary discs \citep{2017ApJTripathi,2018ApJAndrewsA,2020ApJTobinA,2023PPVIIManara} to perform a statistical comparison. For the observational samples, it is best to use systematic surveys with a large sample size, as they offer valuable information about the demographics of exoplanets. This includes:
\begin{itemize}
	\item the fractions of stars with detected planets and the multiplicity of planetary systems;
	\item the distributions of planet properties, such as their masses, periods, and eccentricities;
	\item correlations between these properties; and
	\item correlations between stellar and planet properties, such as the increasing frequency of giant planets with stellar metallicity \citep{1997MNRASGonzalez,2004A&ASantos,2005ApJFischer} or the role of stellar mass and age.
\end{itemize}

For such an undertaking, one can perform planetary population synthesis \citep{2004ApJIda1,2009A&AMordasinia,2014PPVIBenz,2018BookMordasini,2023EPJPEmsenhuber} using a global numerical model. This requires several ingredients: 1) initial conditions that are representative of protoplanetary discs, 2) a global formation and evolution model that is able to predict final planetary system properties from the initial conditions, 3) an observational sample, and 4) a synthetic detection bias to mimic the detection capabilities of the survey in order to qualitatively compare the synthetic and observed populations.

In \citet[hereafter \paperone]{2021AAEmsenhuberA} we presented the Generation III Bern model of planetary formation and evolution. This model includes disc evolution concurrently with accretion of both solids and gas by the protoplanets, planetary interior structures, gas-driven migration, and dynamical interactions. In addition, the model is able to predict many directly observable planetary properties (such as orbital elements, mass, radius, or magnitudes) from giant planets to compact terrestrial-planet systems, provided the initial number of embryos is sufficient.

Then in \citet[hereafter \papertwo]{2021AAEmsenhuberB} we discussed the initial conditions for the model so that they are representative of known protoplanetary discs and compared different populations where we varied the initial number of embryos per system, one of the free parameters of the model. Some simple qualitative comparisons with the detected planets were performed, which revealed that migration is likely too efficient in the model. However, testing of these populations in a rigorous quantitative manner against the known exoplanets is yet to be conducted.

The ultimate goal of these synthetic populations is to compare them with systematic surveys to determine if the characteristics of the observed exoplanets population can be reproduced. Mismatches should lead to changes in the model or its parameters to provide better agreements \citep{2015IJAsBMordasini,2018BookMordasini} and ultimately improve our understanding of planet formation. Synthetic populations that agree well with the observed planets can then, at least in some aspects, be used to make predictions about the not-yet-observed parts of the population.

Such a comparison was already performed in \citet{2009A&AMordasinib} for the first generation of the Bern model, but results were limited by the number of exoplanets known at that time. In \citet{2019ApJMulders}, an earlier version of this model was compared with the Kepler survey results \citep{2018ApJSThompson}. The outcome was that the synthetic and observed populations were similar in multiplicity and inclinations, for instance, although the synthetic population contained too many planets by a factor of about five.

Here, we aim to compare the populations of \papertwo{} with the radial velocity (RV) technique of planetary detection. We chose to compare with the combined HARPS/Coralie survey, whose results were presented in \citet[][hereafter \citetalias{2011MayorArxiv}]{2011MayorArxiv}.\footnote{This paper has not (yet) been published in a refereed journal. The reason is that at the time, 41 planets used in the statistical analysis had not previously been announced in separate refereed detection papers. In the meantime, almost all planets (except seven out of 169 in total in the updated sample) have gone through this process. As the survey continued between 2011 and 2015, new planets were detected around the stars belonging to our sample. These new planets have been included in the table (appendix~\ref{sec:plans}).} The survey has both a baseline of more than 10 years, which allows the detection of planets at several astronomical units, and a high sensitivity of about \SI{1}{\meter\per\second}. It is a volume-limited sample for a representativity of stars \citep{2000ConfUdry}.

However, the synthetic and observed exoplanet populations cannot be directly compared, and observational biases need to be accounted for when making the comparison. This means that either the synthetic systems need to be treated as if they were observed by the same technique as used for the detection of the actual exoplanetary systems, which means that a synthetic detection bias must be applied to the model systems, or the completeness of the survey must be taken into account to infer the underlying distribution of the observed planets. Here, we take the former approach. It is a key feature of our work that we do not try to de-bias the observed population (and therefore extrapolate to non-detectable or poorly detectable regions) but bias the synthetic population.

The organisation of this paper is as follows. We describe the coupled formation and evolution model, the way we set the initial conditions for our population synthesis, and how we applied an observational bias in Sect.~\ref{sec:methods}. The nominal population presented in \papertwo{} is compared with the HARPS/Coralie survey in Sect.~\ref{sec:nominal}. Then, we investigate in Sect.~\ref{sec:opt} how to adapt some parameters of the formation model to better reproduce the observed planets.
A summary and the conclusions are given in Sect. \ref{sec:conclusion}.

\section{Methods}
\label{sec:methods}

\subsection{Formation and evolution model}

Our combined formation and evolution model, the Generation III Bern model, was extensively described in \paperone{}. We hence provide only a brief overview here. This model derives from the work of \citet{2004A&AAlibert,2005A&AAlibert} for the single-embryo formation (Generation I), \citet{2012A&AMordasiniB} for the long-term evolution part (Generation Ib), and \citet{2013A&AAlibert} for the multi-embryo formation without long-term evolution (Generation II).

Planet formation was modelled for a fixed time interval (either \SI{20}{\mega\year} or \SI{100}{\mega\year}, see Sect.~\ref{sec:methods-pop}), during which gas and planetesimal discs evolve together with the forming planets. The planets accrete from both discs, migrate, and dynamically interact, leading to scatterings, giant impacts, or capture into resonances. After the time elapsed, the evolution stage took place. Here, the thermodynamical evolution (cooling and contraction) of each planet was followed individually to \SI{10}{\giga\year}, along with atmospheric escape and tidal migration.

At its basis, the formation model tracks the evolution of a viscous accretion disc \citep{1952ZNatALust,1974NMRASLyndenBellPringle}, where the turbulent viscosity is provided by the standard $\alpha$ parametrisation \citep{1973A&AShakuraSunyaev}. The disc midplane temperature was estimated following \citet{1994ApJNakamoto}, including as sources stellar irradiation and viscous dissipation. Internal \citep{2001MNRASClarke} and external \citep{2003ApJMatsuyama} photoevaporation were included as well. The solid component was represented by planetesimals, whose dynamical state is governed by the drag from the gas and the stirring produced by the other planetesimals and the planets \citep{2004AJRafikov,2006IcarusChambers,2013A&AFortier}.

The model assumes planets form by core accretion \citep{1974IcarusPerriCameron,1980PThPhMizuno}, by accreting planetesimals in the oligarchic regime \citep{1993IcarusIdaMakino,2002IcarusOhtsuki,2003A&AInaba,2003IcarusThommes,2013A&AFortier}, and by giant impacts (planet-planet collisions). The accretion of gas is first constrained by the ability of the forming planets to radiate away the binding energy released during this process. The envelope mass is then obtained by solving the internal structure equations \citep{1986IcarusBodenheimerPollack}. Once the planet reaches the critical mass as found by solving the structure equations (usually on the order of \SI{10}{\mearth}; \citealp{1982PSSStevenson}), cooling becomes efficient and runaway gas accretion can occur. When this happens, the envelope can no longer remain in equilibrium with the surrounding gas disc and contracts instead \citep{2000IcarusBodenheimer,2012A&AMordasiniB,2012A&AMordasiniC}. In that phase, gas accretion is limited by the supply of the gas disc, while the internal structure equations are used to determine the radius and luminosity.

The protoplanets undergo gas-driven planetary migration. For Type~I migration, we adopted the prescription of \citet{2014MNRASColemanNelson}, while Type~II migration was computed using the fully suppressed prescription of \citet{2014A&ADittkrist}, and the transition between the two regimes was done according to \citet{2006IcarusCrida}. Multiple embryos could form concurrently in each system, and the gravitational interactions were modelled using the \texttt{mercury} \textit{N}-body package \citep{1999MNRASChambers}.

After the formation stage is finished, the model switches to the evolution stage, where the planets were evolved individually to \SI{10}{\giga\year} following a procedure similar to \citet{2012A&AMordasiniB}. This stage mainly tracks the cooling and contraction. It also includes atmospheric escape \citep{2014ApJJin,2018ApJJin} and tidal migration \citep{2008CeMDAFerrazMello,2009ApJJackson,2011AABenitezLlambay}. However, since we are mostly interested in the planet masses in this work, the relevance of the evolution stage is limited compared to planet radii probed in transit surveys, for instance.

\subsection{Population synthesis}
\label{sec:methods-pop}

The procedure to compute a full population was presented in \papertwo, which represents an update from \citet{2009A&AMordasinia}. The model assumes that the systems form around single stars, and we used observational data to constrain the initial conditions of the formation and evolution model. The initial conditions were the disc masses \citep{2018ApJSTychoniec}, disc sizes \citep{2010ApJAndrews}, inner edge of the disc \citep{2017AAVenuti}, protoplanetary disc lifetimes, and the dust-to-gas ratio (from the observed stellar [Fe/H] as described in \citealp{2009A&AMordasinia}, but with a solar reference value of the dust-to-gas ratio of \num{0.0149}, \citealp{2003ApJLodders}, corresponding to [Fe/H]$=0$).

Other model parameters were kept constant in one population, such as the parameter of the turbulent viscosity, $\alpha=\num{2e-3}$ (to reproduce the relationship between disc masses and stellar accretion rates, \citealp{2017ApJMulders,2019AAManara}), and the initial slope of the solids' surface density, which is steeper than the one of the gas disc \citep{2018ApJAnsdell}.

In line with our already-published results, we mainly analyse the nominal synthetic planetary population that was presented in \papertwo, called \texttt{NG76}. This population starts with 100 planetary embryos per disc and was computed for a duration of the formation stage (which includes the tracking of \textit{N}-body interactions) of \SI{20}{\mega\year}. However, to take into account that dynamical instabilities can take longer to occur \citep[e.g.][]{2021AAIzidoro}, we also analyse a population where the formation stage was extended to \SI{100}{\mega\year}. This population, called \texttt{NG76longshot}, was already discussed in \citet{2023EPJPEmsenhuber} and \citet{2024NatAsBurn}. Finally, we also consider in Sect.~\ref{sec:opt} a third optimised population, \texttt{NG192}, for which several model parameters were varied to obtain a better agreement with the HARPS/Coralie observations. The results presented in this work are based on the state of the populations at \SI{5}{\giga\year}.

\begin{figure}
	\centering
	\includegraphics[width=\columnwidth]{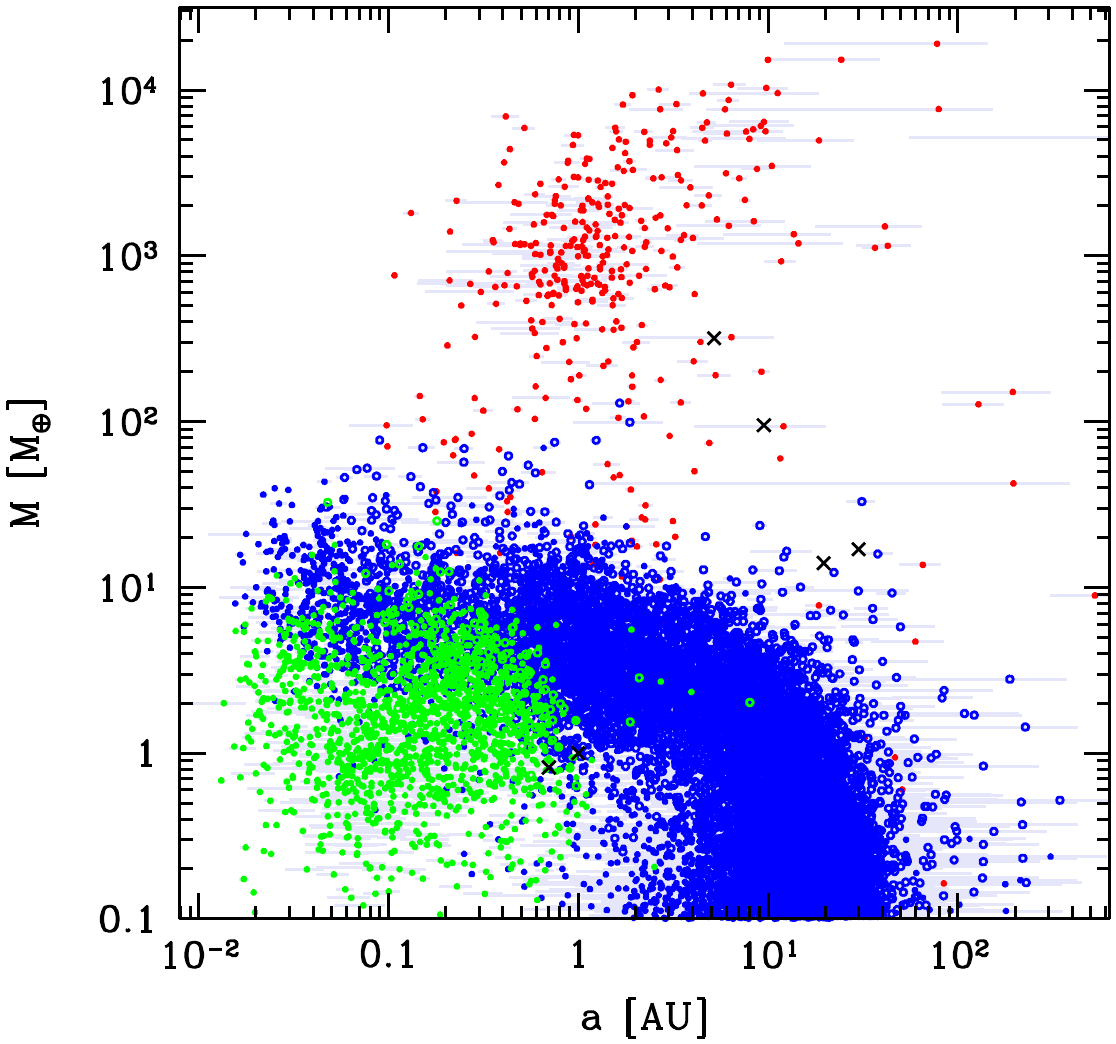}
	\caption{Mass-distance of the (unbiased) \texttt{NG76longshot} population at an age of \SI{5}{\giga\year}. Each point corresponds to one synthetic planet. The population consists of 1000   synthetic planetary systems. Colours represent the bulk composition: red are planet where the H/He mass is larger than the mass of solids (the core mass). Open circles are planets containing H/He but with a core mass higher than the envelope mass. Filled circles correspond to planets without H/He. Additionally, blue (green) indicates that planets have accreted more (less) than 1\% of their core mass as ices. Horizontal bars reach from the peri- to the apoastron to represent eccentricities. Finally, black crosses show the Solar System for comparison.}
	\label{fig:am76longshot}
\end{figure}

\begin{figure}
	\centering
	\includegraphics[width=\columnwidth]{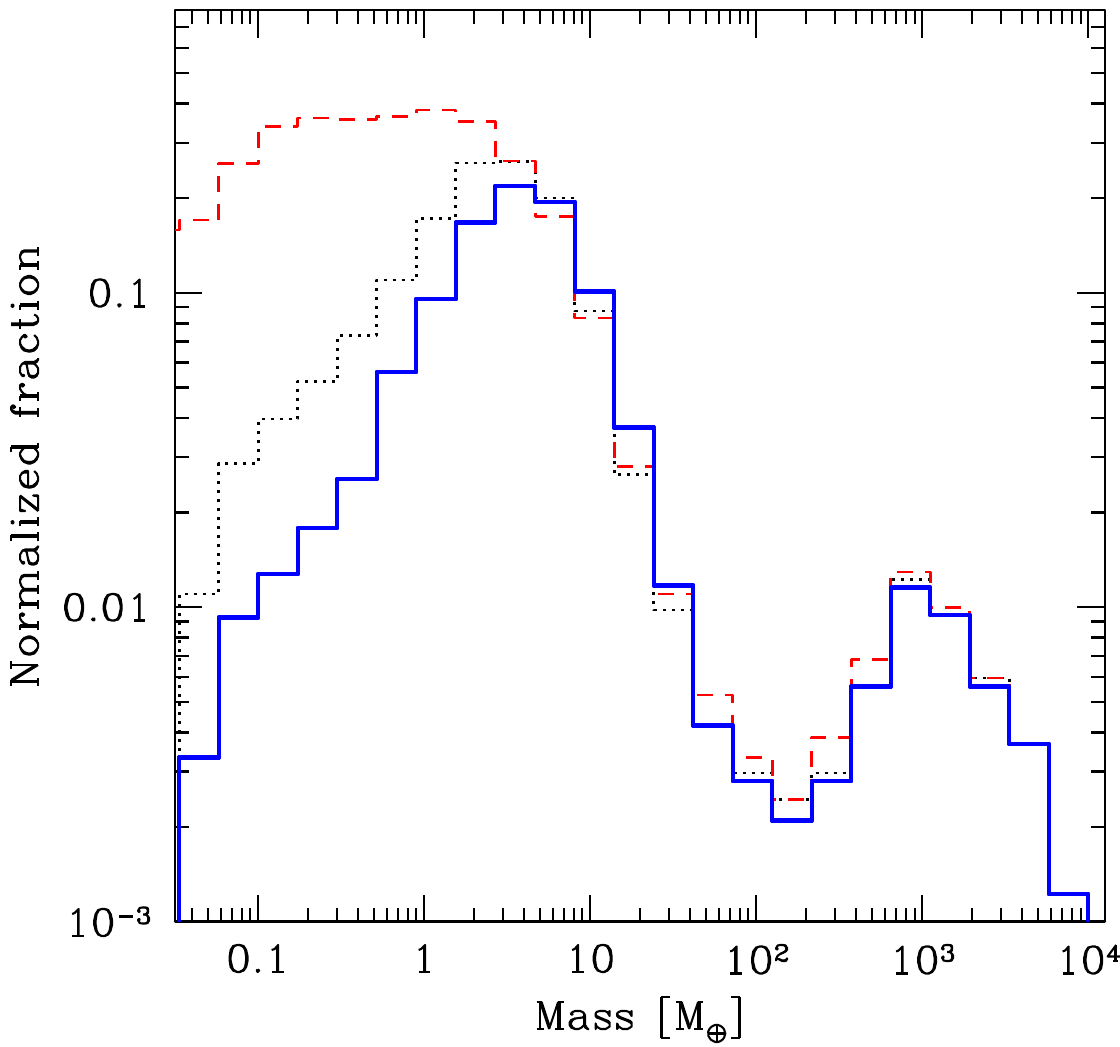}
	\caption{Unbiased mass function for planets within $a<\SI{3}{\au}$ at three moments in time. The blue solid line shows the distribution at \SI{100}{\mega\year} (\texttt{NG76longshot}) while the black dotted one is at \SI{20}{\mega\year} (\texttt{NG76}). The red dashed line additionally shows the situation at the moment when the gas disc disappears in the systems (corresponding to \SI{3}{\mega\year} on average). The reduction of the number of low-mass (proto)planets of mainly (sub-)Mars to Earth mass and the associated increase of more massive super-Earth and (sub-)Neptunian planets is visible. It is a consequence of giant impacts during the post-disc phase.}
	\label{fig:histomass3}
\end{figure}

Figure \ref{fig:am76longshot} shows the (unbiased) mass-distance diagram of the \texttt{NG76longshot}  population. Compared to the situation in \texttt{NG76} at \SI{20}{\mega\year} (see \papertwo), little change of the $a-M$ diagram occurs for Jovian and (sub)-Neptunian planets. However, there is a clear reduction of the number of low-mass planets ($\lesssim \SI{4}{\mearth}$) at orbital distances of about less than \SI{3}{\au}. This is due to numerous giant impacts involving (proto)planets with masses typically between Mars and Earth. These impacts mainly occur in systems consisting only of low-mass planets that have not migrated much (Class I systems according to the planetary system architecture classification of \citealt{2023EPJPEmsenhuber}). This evolution is clearly seen in Fig.~\ref{fig:histomass3} which shows the unbiased planetary mass function at three moment in time: at $\tau_{\rm disc}$ when the gas disc disappears (so 1--\SI{10}{\mega\year} with an average of about \SI{3}{\mega\year}), at \SI{20}{\mega\year}, and at \SI{100}{\mega\year}. One sees the strong reduction of the number of low-mass (mainly sub-Earth) protoplanets occurring mainly between $\tau_{\rm disc}$ and \SI{20}{\mega\year} but continuing also in the following \SI{80}{\mega\year}. The associated mild increase in the number of more massive super-Earth and (sub-)Neptunian planets is also visible. In the giant planet mass range, only minor evolution occurs between $\tau_{\rm disc}$ and \SI{20}{\mega\year}, and even less occurs between 20 and \SI{100}{\mega\year}. However, as becomes clear below, for the subject of this paper which is the statistical comparison of NGPPS and the HARPS/Coralie survey, the changes between 20 and \SI{100}{\mega\year} are of minor importance. The reason is that the survey is hardly sensitive to the very low-mass planets where significant changes occurs.

\subsection{Observational bias}
\label{sec:methods-bias}

To compare our populations, we applied an observational bias using a technique that is updated from \citet{2009A&AMordasinib}. The main changes relative to this work are: 1) accounting for the relative inclinations of the planets with respect to the plane of the disc, 2) an updated map of detection probabilities, and 3) performing multiple observations of the same population to take into account different orientations.

For each observation of the full synthetic population, we randomly selected 822 systems (the same as in the combined HARPS/Coralie sample) without replacement from the population of the \num{1000} systems and applied the following procedure. For each system, we randomly determined the direction of the observer with two angles, $\theta$ and $\phi$, that represent the latitude and longitude of the line of sight with respect the systems' reference frame, which is given by the unit vector $\mathbf{\hat{o}}$. Then, for each planet in the system, we computed the inclination $i$ of the orbit with respect to that vector. For this, we determined the unit vector parallel to its orbital angular momentum $\mathbf{\hat{h}}$ from its inclination to the reference plane $i_\mathrm{p}$ and longitude of the ascending node $\Omega_\mathrm{P}$. The unit vectors are then
\begin{equation}
\mathbf{\hat{o}} = \left(
\begin{array}{c}
\cos{\phi}\sin{\theta} \\
\sin{\phi}\sin{\theta} \\
\cos{\theta}
\end{array}
\right)\ \ \text{and}\ \ \mathbf{\hat{h}} = \left(
\begin{array}{c}
\sin{\Omega_\mathrm{P}}\sin{i_\mathrm{P}} \\
-\cos{\Omega_\mathrm{P}}\sin{i_\mathrm{P}} \\
\cos{i_\mathrm{P}}
\end{array}
\right),
\end{equation}
and the inclination of the orbit with respect to the observer is the angle between the vectors $\mathbf{\hat{o}}$ and $\mathbf{\hat{h}}$. This permitted us to retrieve the $\sin{i}$ value with
\begin{equation}
\cos{i}=\mathbf{\hat{o}}\cdot\mathbf{\hat{h}}\ \ \text{and}\ \ \sin{i}=\sqrt{1-\cos^2{i}}.
\end{equation}
It can be seen that in the case the planet orbits the star in the reference plane (i.e. $i_\mathrm{P}=0$), this method simplifies to the simple case $\sin{i}=|\sin{\theta}|$.

\begin{figure}
	\centering
	\includegraphics{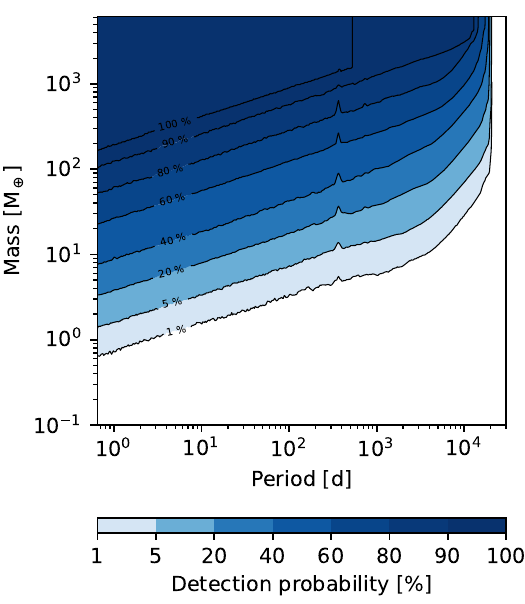}
	\caption{Average probability of planet detection from the HARPS/Coralie survey as function of a planet's period and minimum mass \citep{2011MayorArxiv}.}
	\label{fig:bias}
\end{figure}

Then, it is possible to compute the effective mass of the planet $M\sin{i}$ (for clarity, we here refer to the planet mass as $M$ rather than $\mplanet$ in \papertwo). To retrieve its detection probability, we use the probabilities derived by \citetalias{2011MayorArxiv} for the full HARPS/Coralie survey that depend on $M\sin{i}$ and the orbital period $P$. The detection probability is provided in Fig.~\ref{fig:bias} for reference. This synthetic detection bias was obtained by adding a signal from a planet with the corresponding mass and circular orbit with the prescribed period in the actual observations of the star (with RV signals of detected planet removed) and with different phases, and then checking that the signal is larger than a corresponding \SI{1}{\percent} false alarm probability. Long-period planets with incomplete coverage are treated in the same way. In this work, we just made one adjustment from the detection probabilities: If it is less than \SI{1}{\percent} (the white region in Fig.~\ref{fig:bias}), we unconditionally deemed the planet as not detected. This is to avoid obtaining planets far from the usual detection region of HARPS.

There are a couple of items that should be noted. First, the bias used here was derived from the mean across the stars in the HARPS/Coralie survey. It does not account for the variations between the individual stars and the planetary system architecture. It also does not account for the planet eccentricities. Second, it is possible for massive objects to appear with planetary mass when $\sin{i}\approx0$, a situation that the synthetic populations cannot account for. However, such a scenario is statistically unlikely \citep{2001AAJorissen}, and we therefore do not account for it here.

\subsection{Stars and planets from the HARPS/Coralie survey}
\label{sec:methods-harps}

The comparison dataset is provided by the High-Accuracy Radial Velocity Planetary Searcher (HARPS, \citealt{2003MsngrMayor}) GTO (Guaranteed Time of Observation) survey \citepalias{2011MayorArxiv} which was combined for the statistical results with the Coralie survey \citep{2000ConfUdry}, as described in \citetalias{2011MayorArxiv}. The sample was defined as volume-limited of FGK stars closer than \SI{50}{pc} in the southern sky, according to the distances known at the time. The sample has not been revised based on more recent distance measurements, such as those of GAIA. Known visual (detected companion with 6 arcseconds) and spectroscopic (RV signal indicating a companion larger than \SI{13}{\mj}) binaries were excluded. This leaves a sample comprising \num{822} stars. The mean stellar mass across the whole sample is \SI{0.91}{\msun} with a standard deviation of \SI{0.17}{\msun}. The full distribution of stellar masses is provided in Appendix~\ref{sec:plans}. Concerning the metallicity, the mean value is \num{-0.11} with a standard deviation of \num{0.28}.

The analysis was based on the data from \citepalias{2011MayorArxiv}, with the inclusion of four more years of observations (2011 to 2015). The other items were left as they were used in \citepalias{2011MayorArxiv}. This updated analysis retrieved $\nobs=\num{169}$ planets and candidates around 105 host stars/planetary systems. Detailed parameters about the survey along the list of planets are provided in Appendix~\ref{sec:plans}.

Among the detected planets, 24 of them in 20 systems are in known binary systems. It was found that close binaries result in smaller and shorter lived discs \citep[e.g.][]{2012ApJHarris,2017ApJCox}. Recent studies find that discs become insensitive to binary separation above a distance of about \SI{140}{\au} \citep{2025AARicciardi}. Of the 20 systems, only three system are known to be in a binary closer this limit: HD142 \citep{2016MNRASTokovininKiyaeva}, HD4113 \citep{2018AACheethamA} and HD7449 \citep{2016ApJLRodigas}. Thus, the initial conditions of the synthetic populations, although nominally obtained for single stars, should also be representative of binary systems except the three mentioned here, out of 105. We believe this should have only a small effect on the statistical comparison.

\subsection{Statistical analysis}
\label{sec:methods-stats}

To assess whether the synthetic and observed populations are different, we performed a series of statistical tests. The null hypothesis of these tests is that the observed sample is one random draw from the synthetic population with observational bias applied. For this, we applied the procedure outlined in Sect.~\ref{sec:methods-bias} to obtain \num{1000} samples that can be compared to detected sample.

For single-quantity comparisons, we performed a two-sided Kolmogorov-Smirnov (KS) test between each pair of synthetic and observed sample using the \texttt{SciPy} package \citep{2020NatMeSciPy}. To compute the significance of the test, we followed \citet{1968JRSSBHope} and compute the KS distance at the significance level $S$ that we chose and compared this value to the expected distance from the KS distribution. If the obtained distance is larger than that of the KS distribution, then the null hypothesis is rejected. This is also an improvement relative to \citet{2009A&AMordasinib}, who only computed one observed sample for the single-variable comparison.

For the multi-variable comparisons (such as the mass-period diagram), we proceed in the same way as \citet{2009A&AMordasinib}: we implement the algorithm of \citet{2002BookPress}. We generate \num{1000} samples with 169 synthetic planets each, which are put together to form the comparison population. Then the distance between each of the sample as well as the HARPS/Coralie sample against the comparison population is computed. This enabled us to compute the significance of the 2D KS test directly.

\section{Comparison of the nominal population}
\label{sec:nominal}

\begin{figure*}
	\centering
	\includegraphics{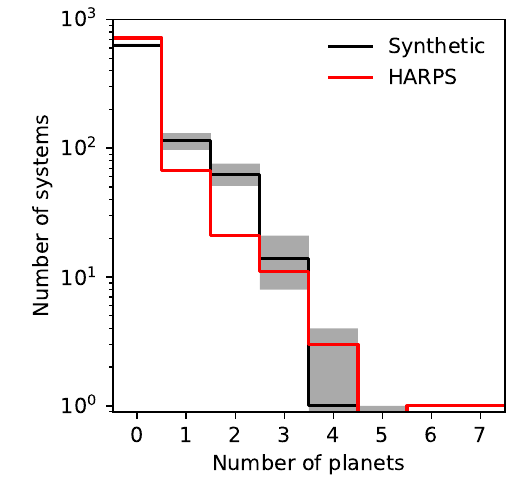}
	\includegraphics{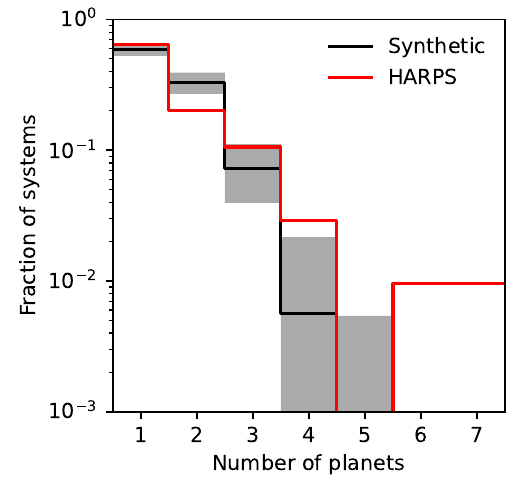}
	\caption{Histograms of the number of detected planets per system in both the HARPS/Coralie surveys (red) and the biased synthetic population (black). The histogram of the synthetic system is based on \num{1000} samples of the synthetic population, with the solid line denoting the median while grey band shows the \SI{95}{\percent} confidence interval. The left panel compares the absolute number of systems, while the right panel compares the relative number of systems that have detected planets.}
	\label{fig:hist}
\end{figure*}

We based our results on two populations from previous works (Sect.~\ref{sec:methods-pop}). First, the \texttt{NG76} population from \papertwo. This is the one with initially 100 lunar-mass embryos per system; the highest number of embryos per system that was considered. Second, we have the \texttt{NG76longshot} population where the formation stage has been extended from \SI{20}{\mega\year} (as in \texttt{NG76}) to \SI{100}{\mega\year} to allow for \textit{N}-body interactions to capture later instabilities.

These synthetic populations were computed for a stellar mass of $\mstar=\SI{1}{\msun}$, which is the most similar value that we have in the synthetic populations to those in the sample (Sect.~\ref{sec:methods-harps}). Populations around lower-mass stars are found in \citet[][NGPPS IV]{2021AABurn}. The distribution of disc metallicities (that we take to be the same as stellar metallicity) is slightly higher than that of the sample, as the mean value is \num{-0.02} \citep{2005AASantos}. Nevertheless, the values are similar enough to provide for a good comparison.

The analysis was performed on \num{1000} Monte Carlo samples of the biased synthetic population, each containing 822 systems, the same number of systems as in the HARPS/Coralie sample. In each sample, the systems were selected without replacement from the \num{1000} systems of the synthetic population. We used the \num{1000} samples to build a distribution and determine the confidence intervals. Table \ref{tab:compnbplanets} summarises the results on the base statistics found in the different populations.

\subsection{Number of detected planets, multiplicity}
\begin{table*}
   \begin{center}
      \caption{Base characteristics of the populations analysed in this work.}
      \label{tab:compnbplanets}
      \begin{tabular}{l c c c c c }
         \hline\hline
         Sample & Type & Comment & Nb of planets & Nb of systems & Mean multiplicity \\
         \hline
         HARPS/Coralie & actual & including candidates  & 169 & 105 & 1.61 \\
         \texttt{NG76} & synthetic & nominal & $290^{+30}_{-28}$ & $193^{+17}_{-16}$ & $1.50^{+0.31}_{-0.25}$\\
         \texttt{NG76longshot}& synthetic  &100 Myr integration time& $294^{+30}_{-27}$ & $200^{+18}_{-17}$ & $1.47^{+0.30}_{-0.25}$\\
         \texttt{NG192} & synthetic & optimised& $179_{-20}^{+22}$ & $120_{-12}^{+13}$ & $1.49^{+0.37}_{-0.29}$\\
         \hline
      \end{tabular}
      \tablefoot{For the three synthetic populations, we give the 95\% confidence intervals. The resulting mean multiplicity is given by the number of planets divided by the number of systems.}
   \end{center}
\end{table*}

To begin with the comparison of absolute numbers, we note that we detect $290^{+30}_{-28}$ planets (\SI{95}{\percent} confidence interval) in our synthetic observed population, compared to 169 planets (including candidates) actually detected in the combined HARPS/Coralie sample. Thus, we have close to the double (factor \num{1.70}) of planets in our synthetic population compared to observations. Comparisons with the Kepler data have already found that populations computed with the Bern model generally produce too many planets. \citet{2019ApJMulders} has shown that populations obtained using an older version of the model had to be scaled down by a factor of about five to match Kepler observations. Analyses performed on \texttt{NG76} by \citet{2023AAMatuszewski} found a similar factor, while \citet[][NGPPS VI]{2021AAMishra} found that, on average, 1.7 planets per star could be detected in the synthetic population compared to an average of 1.32 in Kepler DR25 \citep{2018ApJSThompson}. This corresponds to an overabundance by a factor of roughly 1.3. Thus, both the HARPS (RV) and Kepler (transits) comparisons find an overabundance of planets in \texttt{NG76}, while the actual factor depends on the method used.

We stress here that the model parameters or initial conditions were not selected to match either. The main parameter that controls the efficiency of planetary growth in the current model, the radius of the planetesimals, was selected according to \citet{2013A&AFortier}. The initial conditions were selected to best match disc observation, about their mass, extend, and lifetimes (\papertwo), rather than for the resulting systems. Considering this, we find that the number of planets is in a fairly good agreement with observational data.

In addition, the observed multiplicity is similar in the synthetic and observed populations. This is important, because multiplicity is a proxy for system architecture. The synthetic population has $290^{+30}_{-28}$ planets around $193^{+17}_{-16}$ stars, giving a mean multiplicity of $1.50^{+0.31}_{-0.25}$, while in the observed population there are 169 planets (including candidates) in 105 systems, for a mean multiplicity of \num{1.61}. This means that the synthetic populations has too many systems harbouring each a number of detectable planets similar to the observations, rather than a larger multiplicity in systems with detectable planets. The latter would correspond to different, incompatible system architectures in the model versus the observations.

To support this, we provide in Fig.~\ref{fig:hist} histograms of the number of planets per system. The left panel, which provide an absolute comparison, shows a lack of systems without planets and with large multiplicities, while systems with one, two, or three are all a bit over-represented. The right panels provide a relative comparison of the occurrences of multiplicities in systems with planets, and shows that systems with two planets are the only ones that are over-represented, while systems with three planets or five and more are underrepresented.

As Fig.~\ref{fig:hist} shows, the model is unable to reproduce the systems with five or more close-packed sub-Neptunes (such as HD 10180; \citealp{2011AALovis}). However, this discrepancy should be taken with caution. For one reason, we use the average observational biases. In RV surveys, once a (low-mass) planet (or multiple planets for that matter) has been detected, the system will likely be observed more intensely, potentially revealing planets that could otherwise have been missed. This introduces a human factor in the selection process of the target stars, which is not accounted for here. To determine whether this difference is significant, a more elaborate scheme to mimic the selection process would be required.

\subsection{The planetary mass function}

\begin{figure*}
	\centering
	\includegraphics{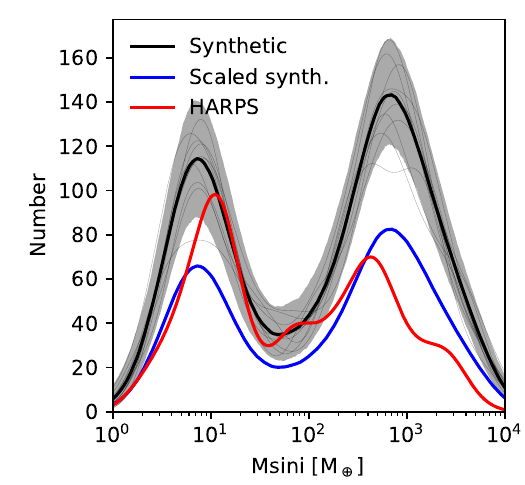}
	\includegraphics{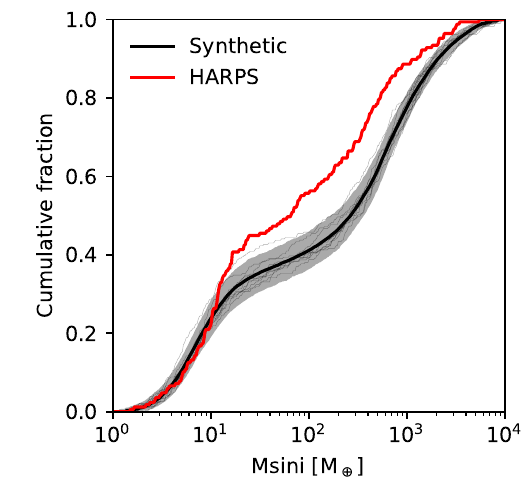}
	\caption{Comparison of the planet masses between the synthetic biased population (\texttt{NG76}, black) and the HARPS/Coralie sample (red). The data of the synthetic population is based on \num{1000} Monte Carlo synthetic observations, with the bold line showing the median of these, the grey region showing the \SI{95}{\percent} confidence interval, and the thin black lines indicating ten individual realisations. \textit{Left}: Kernel density estimate (scaled to be absolute) with a smoothing length of \SI{0.19}{dex}, with the addition of the synthetic population scaled down to the same total number of planets as the observed sample (blue). \textit{Right}: Cumulative distributions of the same data. The statistical analysis shows that the samples from the synthetic population do not match the detected ones at the \SI{5}{\percent} level because the model under-predicts the number of sub-Neptunes and over-predicts the number of giants.}
	\label{fig:mass-comp}
\end{figure*}

The planetary mass function is a central quantity, holding many clues for planet formation models \citep{2004ApJIda1,2009A&AMordasinia,2018ApJSuzuki,2019MNRASNayakshin,2023EPJPEmsenhuber}. It allows for comparison of the relative number of planets of different masses. For this purpose, we show in Fig.~\ref{fig:mass-comp} the comparison of the masses of the \num{1000} biased samples from the synthetic population (in black) and the detected ones (in red). The left panel provides a kernel density estimation (KDE), though it is scaled by the number of planets in each sample for absolute comparison). To compute the bandwidth of the KDE, we perform a search of the best value using \texttt{scikit-learn} \citep{sklearn} on the observed sample, which returns a value of roughly \num{0.19} (in units of decimal exponent). This same bandwidth is then used for all the synthetic observations. As we discussed at the beginning of this section, the number of observable planets in the synthetic population is higher than the number of detected by a factor of roughly two. To provide a relative comparison, the re-scaled synthetic planets to the total number of observed ones in shown in blue. The right panel shows the same data, but in the form of a cumulative distribution. We also perform a two-sample KS test with a rejection criterion for p-value below \SI{5}{\percent} according to the procedure outlined in Sect.~\ref{sec:methods-stats}. We find that the value at the \SI{95}{\percent} of the distribution of KS distances of \num{2.34}. As this value is larger than the limit of \num{1.36}, we therefore rejected the hypothesis that observed HARPS/Coralie planets are a random draw of the synthetic planets with observational bias applied.

The general form of the planet mass distribution with two peaks is, however, clearly retained: one peak at about \SI{10}{\mearth} (Neptunian planets) and one for the giant planets larger than about \SI{300}{\mearth}. Nevertheless, the following differences remain: 1) an overabundance of giants in relation to sub-Neptunes in synthetic planets compared to the HARPS/Coralie data, 2) a larger dip at intermediate masses in the synthetic planets (as was already pointed out by~\citet{2019MNRASNayakshin}, \citet{2021AJBennett}, and also found by \citet{2022AASchlecker} in a low-mass star population from our model (\citealp{2021AABurn}, NGPPS IV), and 3) lower masses for the sub-Neptunes in the synthetic planets.

For the first item, the cumulative distribution on the right panel of Fig.~\ref{fig:mass-comp} reveals the amount of the imbalance between sub-Neptunes and giants. If we take the transition between the two categories at \SI{100}{\mearth}, we obtain the following values of the cumulative distributions: \num{0.41\pm0.05} for the biased synthetic population compared to \num{0.56} in the HARPS/Coralie sample. This means that if the number of sub-Neptunes remains the same, the number of giants would need to be reduced by about \SI{45}{\percent} to balance back the two categories. Alternatively, the number of sub-Neptunes would need to be increased by about \SI{83}{\percent}.

Now considering the depth of the planetary desert, we compute the fraction of planets between \num{20} and \SI{200}{\mearth} (the shallowest part in the cumulative distribution of the synthetic population). We find that in the synthetic population, \SI{14}{\percent} of the planets are in that range while for the HARPS/Coralie sample, that value is \SI{22}{\percent}. This means that the number of planets in the desert would have to be increased by \SI{57}{\percent} (assuming the number of planets outside the desert remains the same) to balance the synthetic population. This is a factor of less than two, which we think is not a strong disagreement with the observations, given the following fact: our model starts with 100 moon-mass embryos per disc and all further evolution is given by the physical processes included in the global model in a low-dimensional description (like an axisymmetric disc or radially symmetric planets)  (\paperone).

We also emphasise that the presence of a planetary desert in the observed population is subject to debate. For instance, \citet{2021AJBennett} suggest that no such desert exists and that the populations presented here have much larger differences on the relative number of planets in the desert than the \SI{57}{\percent} we obtained. Conversely, \citet{2022MNRASBertauxIvanova} find that the planetary desert does exist and their analysis of synthetic populations from an earlier version of the same model \citep{2018BookMordasini} found the desert in the synthetic populations to not be large enough. Our results are consistent with the second work \citep[i.e.][]{2022MNRASBertauxIvanova} in that we do find a planetary desert, even though the depth of the desert is too deep in the synthetic population analysed here. As already discussed in several works \citep{2011AAMordasini,2018ApJSuzuki,2019MNRASNayakshin}, the too deep desert suggests that our model overestimates gas accretion rates of intermediate mass planets. In this mass regime, planets change from the planet-limited (or attached phase) to disc-limited (or detached phase) gas accretion.

The offset in stellar masses between the synthetic (where it is fixed to $M_\star=\SI{1}{\msun}$) and the CORALIE/HARPS sample (mean of about $\SI{0.91}{\msun}$) might explain a part of the discrepancy. \citet{2021AABurn} analysed the effect of the stellar mass on synthetic populations from the same model. They found that decreasing the stellar mass tend to decrease the number of sub-Neptunes and giants, though the latter are more affected. A synthetic population with a lower stellar mass would reduce the number of planets overall and the imbalance between sub-Neptunes and giants. However, this effect would not be sufficient to resolve the discrepancy entirely.

For the location of the peaks, we note that the synthetic population exhibits a peak centred slightly below \SI{1000}{\mearth} while the HARPS/Coralie planets peak is located closer to \SI{500}{\mearth}. These planets are dominated by their envelopes. Thus, the envelope mass of the planets in the synthetic population is too large. This points to disc-limited gas accretion rates that are overestimated for planets in this mass range. The aforementioned lack of planets in the intermediate regime points towards the same conclusion. There are several possible explanations: differences in the accretion rates obtained in 1D versus 3D simulations \citep{2012MNRASAyliffeBate,2022AASchib}, higher dust opacities than assumed here \citep{2014AAMordasiniB,2019AASchulik}, or a quenching of the gas accretion rate because of more efficient gap opening in low-viscosity discs \citep[e.g.][]{2023ApJAoyamaBai}. To explore the latter point, we have recently added a MHD-wind driven disc model in the Bern model \citep{2023AAWeder}. Future work will show whether this will result in a less dry desert that is in better agreement with observations (see also \citealt{2018ApJSuzuki}).

For sub-Neptunes, however, the mass difference has to be caused by the accretion of solids, as these planets are core dominated. As it can be seen in the right panel of Fig.~\ref{fig:pop-mp-bias}, the synthetic planets are ice-rich for the majority. These planets have formed outside the ice line and subsequently migrated to their final location without accreting much material at the same time. Planets whose masses are about \SI{10}{\mearth} cannot grow from a static solid disc in-situ at the location they are observed (inside about \SI{400}{\day} or \SI{1}{\au}) because there is not enough mass available. The masses of the sub-Neptunes is linked to the behaviour of gas-driven migration. In this range, migration becomes more efficient as mass increases. The planet masses thus reflect the equality mass \citep{2023EPJPEmsenhuber} at which the migration timescale becomes smaller than the accretion timescale. The sub-Neptunes end near the inner edge of the disc, where further growth is nearly absent . If the actual sub-Neptunes indeed follow a similar formation track, then the lower masses of the sub-Neptunes in the synthetic planets relative to the HARPS observations indicates that equality mass is underestimated in the model, which in turn observationally constrains the ratio of the migration to the accretion timescale. Here, the difference indicates a shorter growth and/or longer migration timescale than in the model.

\subsection{Periods}

\begin{figure}
	\centering
	\includegraphics{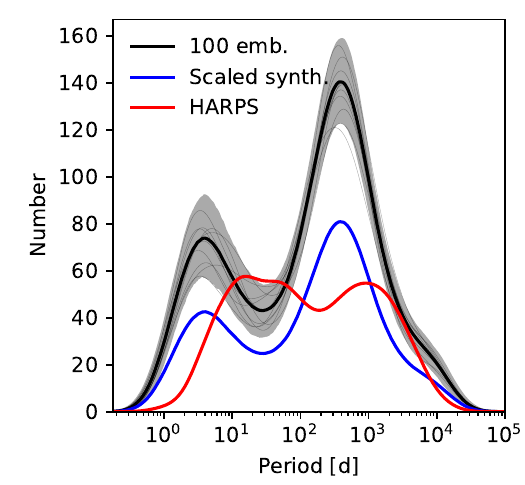}
	\caption{Comparison of the planet periods between the biased synthetic population (\texttt{NG76}, black and blue lines) and the observed HARPS/Coralie sample. The curves have the same meaning as in the left panel of Fig.~\ref{fig:mass-comp}.}
	\label{fig:period-comp}
\end{figure}

The other main quantity retrieved from RV surveys is the orbital period. To check for how the synthetic population and the Coralie-HARPS sample, we proceeded as for the minimum mass. The comparison of the KDEs is provided in Fig.~\ref{fig:period-comp}.

The result shows that the two populations again exhibit similar features, with the two peaks for the sub-Neptunes and giant planets. Giant planets are overrepresented in the synthetic populations, as discussed previously.

The match between the periods in the synthetic and observed populations is not as good as for the masses. For instance, the planets in the synthetic population are found at shorter periods than the observed planets. This is true for both the sub-Neptunes (between \num{1} and \SI{30}{\day} versus \num{3} and \SI{100}{\day} for the observed planets) and giants (between \num{100} and \SI{1000}{\day} versus \num{300} and \SI{3000}{\day}). Such a difference was already seen in the unbiased population, as discussed in \papertwo. This inward shift of the low-mass planets is consistent with \citet{2019ApJMulders}, who found the same feature in the comparison with the Kepler planets. Further, the peaks in the period distribution are narrower in the synthetic population.

For the sub-Neptunes, their final location is near the inner edge of the gas disc. This means that either the choice of the disc inner edges is incorrect or the sub-Neptunes stop migrating before reaching the inner edge. In the synthetic populations, the disc inner edges were assumed to be at the corotation radius stellar rotation and used a rotation period distribution of young stellar objects \citet{2017AAVenuti}. As discussed in \papertwo{}, this provides good agreement with gas tracers; we thus do not think this is the reason for the discrepancy. There are, however, proposed mechanisms to halt the inward migration outside of the inner edge. For instance, there could be viscosity transitions which act as migration traps, such as near the silicate evaporation line, as suggested by \citet{2019AAFlock}. We shall investigate this in future work.

For the giant planets, their locations change only marginally once they undergo runaway gas accretion. Thus, the difference in their final location is most likely due to the migration when planets are at an intermediate mass range (\SI{\sim10}{\mearth}) or the planets originate further out than what our model predicts. We investigate the effect of migration strength in Sect.~\ref{sec:opt}.

\subsection{Mass-period diagram}
\label{sec:nominal-dist}

\begin{figure*}
	\centering
	\includegraphics{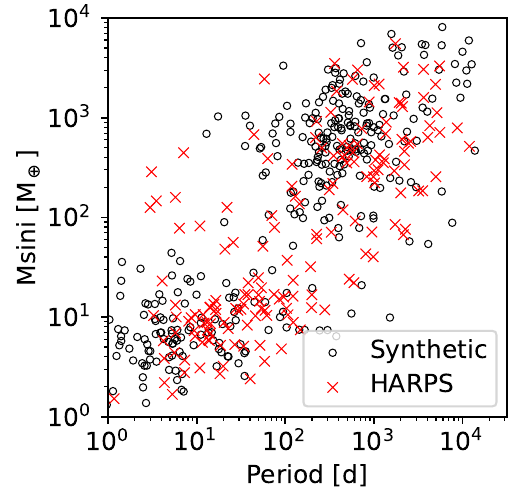}
	\includegraphics{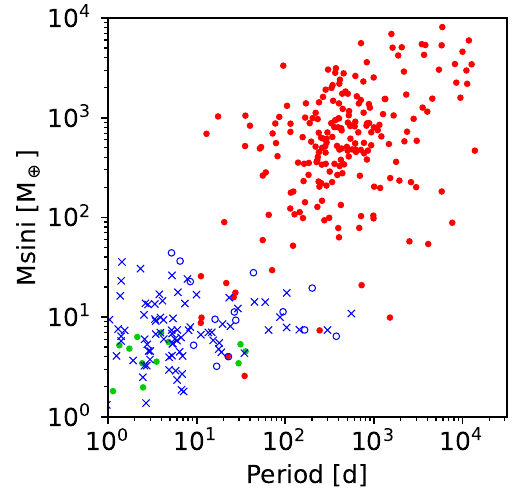}
	\caption{\textit{Left}: Comparison of the minimum mass ($M\sin{i}$) versus orbital period of the biased synthetic 100-embryo population (\texttt{NG76}, black circles) and the actual HARPS/Coralie survey \citepalias[red crosses;][]{2011MayorArxiv}. Both samples have 822 systems. \textit{Right}: Same but only for the synthetic population and with different colours and symbols depending on planet properties. Red circles indicate planets whose H/He envelope content is larger than the core; otherwise, green circles are for planets whose cores do not contain any ices, and blue is for the planets whose cores contain ices. In the latter case, circles are for planets whose envelope mass is at least \SI{10}{\percent} that of the core, and crosses are for the others.}
	\label{fig:pop-mp-bias}
\end{figure*}

The main quantities retrieved from planets detected with the RV technique are the minimum mass and the orbital period. We thus want to check whether our model can reproduce both these quantities at the same time. A comparison of these two properties between one random observation of the synthetic population and the HARPS planets is shown in Fig.~\ref{fig:pop-mp-bias}. We also perform the 2D KS testing procedure and find that its significance is \num{<0.001}, meaning that the KS distance of the observed sample is larger than any of the synthetic samples.

From Fig.~\ref{fig:pop-mp-bias} we observe that both synthetic and observed populations have the same similar shape with a group of sub-Neptunes at short orbital period (between \num{2} and \SI{30}{\mearth} and inside of \SI{100}{\day}) and a second group of more distant giant planets (above \SI{100}{\mearth} and \SI{100}{\day}). However, there are several differences, which explain the difference that is highlighted by the KS test.

First, the positions of the planets in the synthetic population are more grouped than that of the observed planets.
The formation of distinct clusters of planets -- despite the stochasticity introduced by \textit{N}-body interactions -- is a general feature of our model.
These clusters form early and trace different growth and migration tracks during their formation phases (see \citealt{2021AASchleckerB} (NGPPS V) and \citealt{2023EPJPEmsenhuber}).

Second, the synthetic population does not cover the whole parameter space of the observed planets.
For instance, the synthetic population does not contain any hot-Jupiter planet (within \SI{10}{\day} and above \SI{100}{\mearth}). This effect is not due to the sampling of the underlying population, but the population itself does not contain any such planet (\papertwo). A possible reason for the mismatch is that the model only considers disc migration as formation channel, while other channels, such as eccentric migration \citep{2018ARAADawson}, are not included. In the latter case, it is because the model does not include tidal circularisation. As an estimate of the number of hot-Jupiters that could potentially form by this channel, we can take planets larger than $\SI{100}{\mearth}$ that have collided with their star. We find that 20 such planets, of which had an inclination of more than \SI{45}{\degree} shortly before impact. For comparison, we would expect five to ten hot-Jupiters in our 1000-star population (hot-Jupiters have an occurrence rate of \num{0.5}-\SI{1}{\percent}, \citetalias{2011MayorArxiv}, \citealt{2012ApJWright}). Thus, it is likely the model could form hot-Jupiters by eccentric migration if all the relevant physical processes were included in it.

\subsection{Correlation with stellar metallicity}
\label{sec:feh}

\begin{figure}
	\centering
	\includegraphics{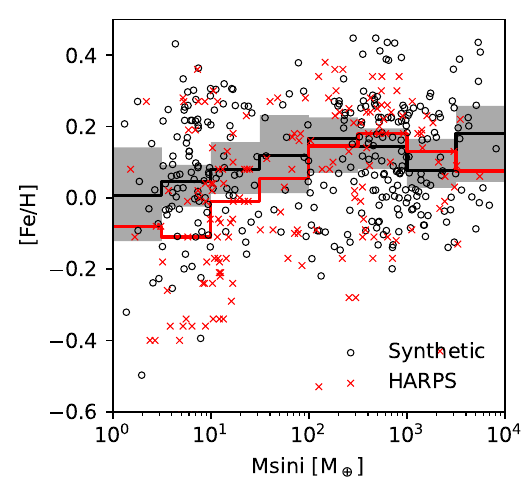}
	\caption{Stellar metallicity versus planet mass ($M\sin{i}$) for the biased 100-embryo population (\texttt{NG76}, in black) and the HARPS/Coralie survey \citepalias[in red]{2011MayorArxiv}. The scatter plot shows one random observation of the synthetic population against the HARPS/Coralie planets. The lines show the medians in the respective range. For the synthetic planets, the values are obtained over \num{1000} simulated observations of the underlying population and showing the \SI{95}{\percent} confidence interval.}
	\label{fig:pop-fm-bias}
\end{figure}

Giant planets are preferentially found around metal-rich stars \citep[e.g.][]{2001AASantosA,2005ApJFischer,2019GeoSciAdibekyan}, while such a correlation is weaker or non-existent for low-mass planets \citep[e.g.][]{2018AJPetigura}. To determine if we recover the same correlation and the same strength, we show a planetary mass versus stellar metallicity diagram for both the combined HARPS/Coralie sample and synthetic population in Fig.~\ref{fig:pop-fm-bias}. In addition, to highlight possible trends in the data, we show the medians for various planet mass bins. As before, the grey zone shows the \SI{95}{\percent} confidence interval for the medians in each bin.

The HARPS/Coralie planets show a positive correlation between mass and stellar metallicity for planets until about \SI{e3}{\mearth} and slightly decreasing afterwards, indicating that the giant planets have a larger metallicity effect that the sub-Neptunes. The synthetic planets show a similar, albeit shallower, correlation; there are only two out of eight bins that do not fit the \SI{95}{\percent} confidence internal.

The two bins that do not fit in the confidence intervals are for sub-Neptune and their median metallicities are too large compared to the HARPS planets. Looking at the individual planets reveals that the synthetic population has an overabundance of sub-Neptunes around high-metallicity stars, while it is missing planets at low stellar metallicities. For instance, the synthetic population cannot reproduce the few systems of sub-Neptunes at metallicities between \num{-0.3} and \num{-0.4}. The systems (HD40307 and HD136352) have a significant mass in planets (about \SI{23}{\mearth} in both cases) for such low metallicities, which indicates that either conversion from dust to planets is efficient or the protoplanetary discs were more massive than what our methodology provides for initial conditions.

Concerning the giant planets, previous studies that used older versions of our formation model, such as \citet{2009A&AMordasinib}, already found lower metallicity effects than in observations. Thus, the shallower metallicity effect that we observe here is not a new feature of our model.

There are multiple possible reasons for the differences between the synthetic population and observations. We may cite a non-linear relation between the metallicity and the usable mass for planetary growth arising during the early phases of planetary growth not considered here \citep{2005ApJYoudinGoodman,2019ApJLenz} or the impact of envelope enrichment \citep{2016AAVenturini} which is neither included in the model. Also, the gas disc properties are assumed to be independent of the metallicity. Finally, metal-poor stars tend to exhibit a higher relative abundance of $\alpha$-elements \citep{2013AAAdibekyan,2017AADelgadoMena}, meaning that they are not as metal-poor as indicated by their [Fe/H] only. Thus, using [Fe/H] with scaled solar composition -- as we do in the model -- results in a mismatch in the total amount of heavy elements, mostly underestimating the total amount of solids for low-[Fe/H] stars. This is of importance, as the total mass of solids is a key quantity in determining the resulting planetary systems \citep{2023EPJPEmsenhuber}.

\subsection{Period ratios of adjacent planets}
\label{sec:pratio}

\begin{figure*}
	\centering
	\includegraphics{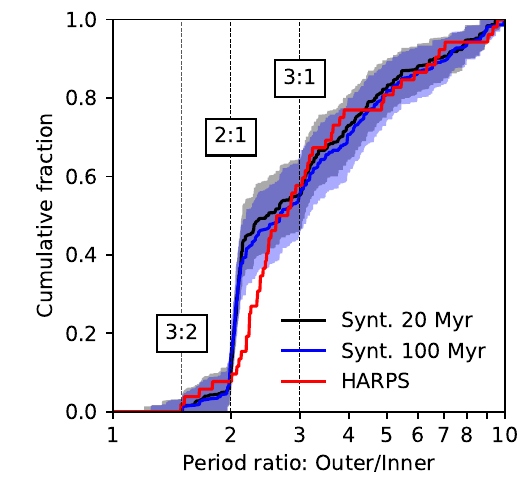}
	\includegraphics{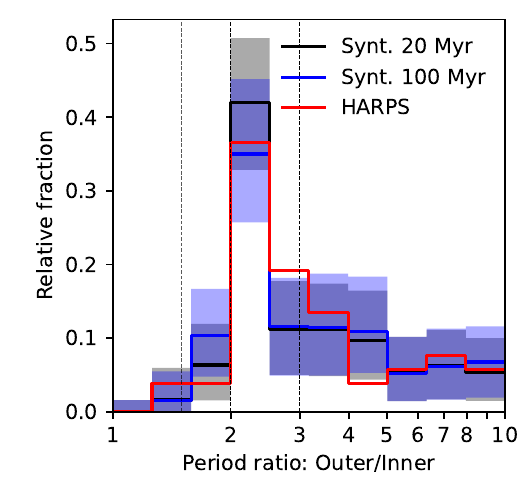}
	\caption{Cumulative distribution (left) and histogram (right) of the period ratios of adjacent planets for the biased 100-embryo population (\texttt{NG76}, in black), the same population but with \textit{N}-body interactions extended to \SI{100}{\mega\year} (\texttt{NG76longshot}, in blue), and the HARPS/Coralie surveys \citepalias[in red]{2011MayorArxiv}. For the synthetic planets, the values are obtained over \num{1000} simulated observations of the underlying population and showing the \SI{95}{\percent} confidence interval. The number of observable pairs is limited, as only systems with multiple planets are usable. The locations of several MMRs are highlighted.}
	\label{fig:comp-pratio}
\end{figure*}

For systems with multiple planets, the spacing between the adjacent planets can be quantified by the ratio of their orbital periods. Figure~\ref{fig:comp-pratio} provides the comparison for this quantity between our synthetic population and the HARPS/Coralie sample. We stress that only systems with multiple detected planets can be used to compute period ratios and that we do not include any bias for the planet orbital periods. As such, the distribution is based on a limited number of planet pairs.

The general outcome of the comparison is that the two curves are in good agreement, with the exception of the pile-up of period ratios past the location of the 2:1. As shown in the histogram (the right panel of Fig.~\ref{fig:comp-pratio}), the number of planets in this pile-up is similar in both synthetic and observed populations. However, the period ratios occupy a much narrower in the synthetic population than in the HARPS/Coralie planets (up to about 2.5). Concerning the other resonances, the synthetic population shows other pile-ups of planets near the 3:1 and 4:1 resonances, though in this case, there is no large discrepancy in the fraction of planet pairs located there between the synthetic and observed systems.

Convergent gas-driven migration will lead to pile-ups at the location of mean-motion resonances (MMRs). In previous populations that were generated with the older version of our formation model (e.g. \citealt{2011A&AAlibert} and \citealt{2019ApJMulders}), the number of pairs close to MMRs was found to be too large compared to the observed exoplanets \citep[e.g.][for results from the Kepler observatory]{2014ApJFabrycky}. Further integration past the dispersal of the gas disc reduces the number of planet pairs locked in MMRs (see Fig.~15 of \citealp{2023PPVIIWeiss}), which is what we are doing in this work.

The standard model \texttt{NG76} follows the dynamical interactions up to \SI{20}{\mega\year} (\paperone). We also performed the same calculations but where the formation stage of the model, during which the \textit{N}-body interactions are simulated, has been extended to \SI{100}{\mega\year} (\texttt{NG76longshot}). This allows us to determine if dynamical interactions at later times would be able to break the resonances. As can be seen in Fig.~\ref{fig:comp-pratio}, we find that this is not the case, and the cumulative distribution of the period ratios is very similar. This means that the \textit{N}-body interactions after \SI{20}{\mega\year} are unable to break the resonances in the period regime accessible to the RV technique. This result is the opposite of that of \citet{2023PPVIIWeiss}, who analysed the same population, but from the perspective of transit observations. As RV surveys are only sensitive to more massive planets, they are less affected by long-term dynamical effects (\papertwo). Thus, the different results between our study and that of \citet{2023PPVIIWeiss} are not contradictory.

This means that we are left with the possibilities that there are other sources of excitation not taken into account in our model, or that fewer planets start off locked in resonances in the first place. This could be the case if migration is reduced overall, as might be expect for MHD-winds driven discs \citep{2010ApJSuzuki,2018AAOgihara} compared to classical $\alpha$-discs as used here. Another possibility is the strength of eccentricity damping that was found to affect the proximity of the planet pairs from MMRs \citep{2022MNRASCharalambous}.

\subsection{Eccentricity}
\label{sec:ecc}

\subsubsection{Overall distribution}

\begin{figure*}
	\centering
	\includegraphics{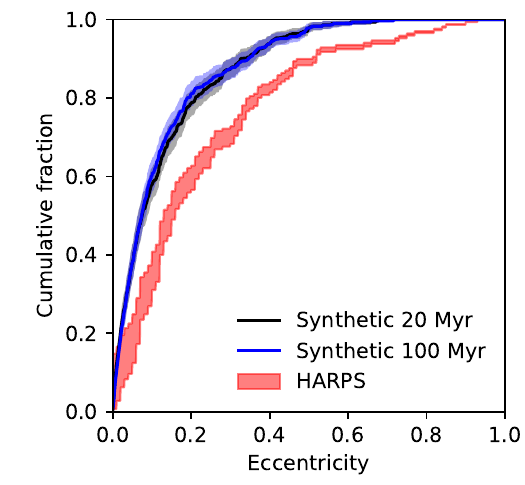}
	\includegraphics{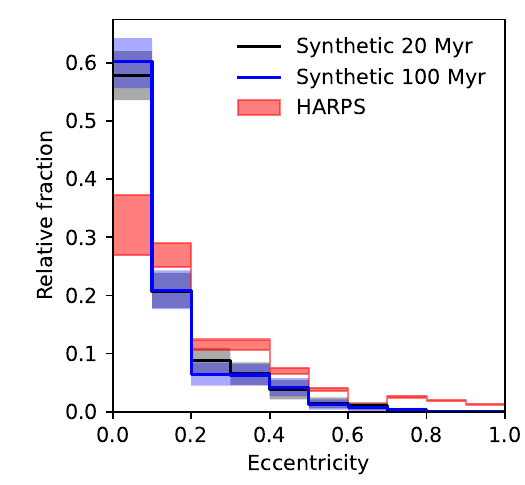}
	\caption{Cumulative distribution (left) and histogram (right) of the planets' eccentricities for the biased 100-embryo population (\texttt{NG76}, in black), the same population but with \textit{N}-body interactions extended to \SI{100}{\mega\year} (\texttt{NG76longshot}, in blue), and the HARPS/Coralie surveys \citepalias[in red]{2011MayorArxiv}. For the synthetic planets, the values are obtained over \num{1000} simulated observations of the underlying population and showing the \SI{95}{\percent} confidence interval.}
	\label{fig:ecc-cumul}
\end{figure*}

Exoplanets have overall higher eccentricities than the planets of the Solar System \citep[e.g.][]{2007ARAAUdrySantos}. Gravitational interactions between the proto-planets stir their eccentricities, and close encounters (or scattering) can lead to very eccentric orbits, up to unity \citep{2019AACarrera}. Scatterings are able to reproduce the eccentricity distribution of the observed exoplanets \citep{2008ApJChatterjee,2008ApJJuric,2008ApJFordRasio}.

To compare the eccentricities between the synthetic population and the observation data, we provide in Fig.~\ref{fig:ecc-cumul} the cumulative distributions of the two. For the HARPS/Coralie sample, we draw two curves: one where undetermined eccentricities are neglected and upper limits are taken at those value (which gives an upper boundary) and one where eccentricities in the same situations are taken as zero (which gives a lower limit). The general form of the two curves is reasonably similar. However, we find a difference in the near-zero eccentricities. We observe that many of the synthetic planets have low eccentricities, with a median value of \num{0.07}. While the uncertainties in the HARPS/Coralie surveys can account for a certain percentage of low-eccentricity planets, it cannot account for all the amount there is in the synthetic population. As the histogram shows, the bin that contains the largest number of observed planets is between $0.1$ and $0.2$, while in the synthetic population, it is for eccentricities below $0.1$, which contains nearly \SI{60}{\percent} of the planets. Conversely, the eccentricities of the observed planets almost reach unity, with the largest value being \num{0.93} for HD20782b \citep{2006MNRASJones}. This is not the case for the synthetic population where the largest value is about \num{0.6}. Our populations do have planets with large eccentricities; they are, however, further away than what is detectable by HARPS.

The discrepancy may come from observations, the model, or both. For RV observations, it is well known \citep{1971AJLucySweeney} that when the eccentricities are small and the S/N is not excellent, planet eccentricities are easily overestimated \citep{2019MNRASHara}. For instance, \citet{2011MNRASZakamska} found in their more sophisticated analysis that about \SI{38\pm9}{\percent} of exoplanets detected by RV have $e<0.05$, compared to \SI{17}{\percent} in standard analysis. The former value is closer to our synthetic population. Comparison with Kepler data also hints that eccentricities in RV could be overestimated. For instance, \citet{2023PPVIIWeiss} find that planets in multiple systems as found by the Kepler survey have eccentricities of \num{0.03} to \num{0.05}, which is even lower than our synthetic populations. Another item is that in RV surveys, it is impossible to distinguish a pair of circular planets on a 2:1 MMR from a single, eccentric planet. This means that some planet pairs could be misinterpreted as high-eccentric planets. However, we note that the planet pairs close to the 2:1 MMR are similar in the synthetic and observed planets (Fig.~\ref{fig:comp-pratio}, right panel), so we do not believe this affects many systems.
Concerning the models, the only source of eccentricity in our simulations is gravitational interaction between the protoplanets. Planet-disc interactions are another potential source, such as for high-mass planets \citep{2006AAKleyDirksen} and turbulent stirring for low-mass planets \citep{2012ApJOrmelKobayashi,2016ApJKobayashi}, but we do not include these effects in our model. Hence, as for the period ratio, we perform the analysis for the population \texttt{NG76longshot} where the formation stage has been extended to \SI{100}{\mega\year} to allow for late dynamical instabilities to occur \citep[e.g.][]{2021AAIzidoro}. We find that the eccentricity distribution between the two populations is very similar. This result is consistent with our findings from the period ratios in the previous section in that later dynamical interactions have a limited effects on the system's dynamical state. Again, this indicates that eccentricities should either be stirred at early times, for instance with a less efficient damping by the gas disc, or that another (potentially external) process currently not included in our model is responsible for pumping the eccentricities. This could be stellar encounters \citep{2011MNRASMalmberg,2020MNRASStock} or the Kozai-Lidov mechanism caused by an external massive companion. Support for the latter hypothesis comes from the fact that many of the large-eccentricity planets in the sample are actually in binary systems, such as HD20782 \citep{2007AADesideraBarbieri}, HD4113 \citep{2018AACheethamA}, HD156846 \citep{2008AATamuz} and HD7449 \citep{2016ApJLRodigas}.

\subsubsection{Mass-eccentricity relation}

\begin{figure}
	\centering
	\includegraphics{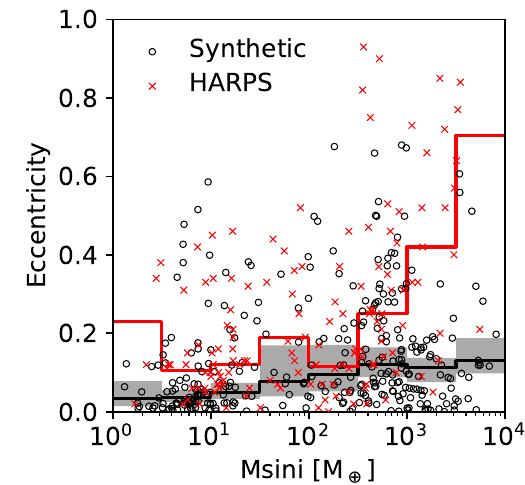}
	\caption{Eccentricity as a function of planet mass ($M\sin{i}$) for the biased 100-embryo population (\texttt{NG76}, in black) and the HARPS/Coralie surveys \citepalias[in red]{2011MayorArxiv}. The scatter plot shows one random observation of the synthetic population against the HARPS/Coralie planets. The lines show the medians in the respective range, but one should note the low number of planets in the bins on the right. For the synthetic planets, the values are obtained over \num{1000} simulated observations of the underlying population and showing the \SI{95}{\percent} confidence interval.}
	\label{fig:ecc-scatter}
\end{figure}

In observational results, higher mass planets tend to have higher eccentricities \citep{2009ApJWright,2011MayorArxiv}. To compare our results with these findings, we provide Fig.~\ref{fig:ecc-scatter}, which is a scatter plot of the eccentricity as a function of mass, along with a running median for each population.

The median of the HARPS/Coralie planets shows two different regions. For intermediate-mass planets (less than about \SI{300}{\mearth}), the median eccentricity is relatively constant with a value of roughly 0.1. For larger masses, however, the median eccentricity increases with the mass, reaching a value larger than 0.6 at the Deuterium burning limit. The synthetic planets, however, do not exhibit a change of behaviour at \SI{300}{\mearth}. Rather, the median eccentricity generally increases with the mass. This is particularly striking at masses \SI{100}{\mearth} where many of the synthetic planets have eccentricities on the order of \num{0.1}, whereas many of the HARPS/Coralie planets have eccentricities of \num{0.4} or more.

As we noted above, the synthetic population is lacking planets with eccentricities larger than 0.6. Thus, a median eccentricity comparable to that of the HARPS/Coralie planets for the largest masses cannot be achieved in the first place. Thus, the giant planets in the synthetic population are dynamically cold compared to the HARPS/Coralie planets. There are synthetic planets with very large eccentricities in the underlying synthetic population, but they are starting just further out than the detection limit of HARPS, which is about \SI{15}{\au} (period of about \SI{2e4}{\day}). The implication is that the model is unable to produce eccentric giant planets without increasing their semi-major axis as well (as the giant planets mainly form within \SI{10}{\au}). Meanwhile, there are several cases of planets with $e>0.8$ and periods less than \SI{e3}{\day} in the HARPS sample.

\section{Towards a better match of observational data}
\label{sec:opt}

In the population synthesis scheme \citep{2015IJAsBMordasini,2018BookMordasini,2018ApJChambers}, one aim is to find model parameters that result in a population that best reproduce the observed exoplanets. In this section, we build on the analysis of the nominal population to perform modifications to the model that result in better agreement with the HARPS/Coralie data.

Although the model has a significant number of parameters, there are none that affects one aspect only of the populations. For instance, one of the main quantities affecting the accretion rate of the planets is the planetesimals' sizes. Increasing their sizes decreases the solids accretion rate of the embryos, thereby leaving a lower number of cores reaching a large enough mass to undergo gas runaway accretion before the dispersal of the gas disc. However, this is not the only effect of the planetesimals size. When planets take longer to form, their migration pattern will also be different. For example, a lower solid accretion rate results in a lower equality mass (see \citet{2023EPJPEmsenhuber} for the definition), which is the mass where the solid accretion and the migration timescales are equal. At this mass, the tracks in the $a-M$ diagram bend inward, and often rapid Type I migration sets in. This might cause the protoplanet end up as a close-in (sub-)Neptunian planet, and not just as a lower-mass, but still distant, planet.

To better reproduce the HARPS/Coralie results, we computed several new populations, using only 20 embryos per system. This is different than the canonical population---which has 100 embryos per system---to limit the computational requirements. We verify in Appendix~\ref{sec:nemb} that this modification does not affect the planetary mass function. This is because the number of embryos affects mostly the Earth-mass planets, while it only has limited effects on the most massive planets (\papertwo). We therefore anticipate that the change of the number of embryos will not affect the results presented here.

\begin{figure}
	\centering
	\includegraphics{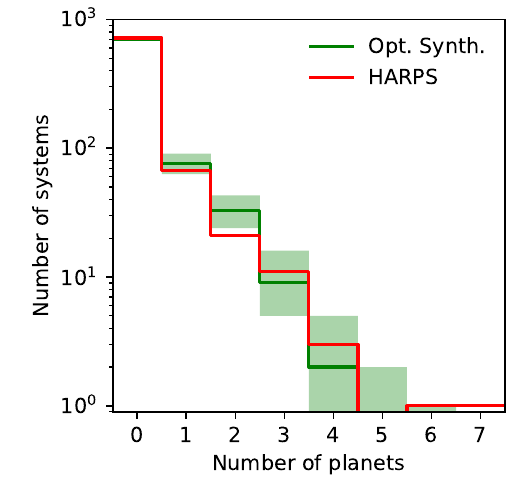}
	\caption{Histogram of the number of planets per systems in both the HARPS/Coralie surveys (red) and the optimised synthetic population \texttt{NG192} (green).}
	\label{fig:match-mult}
\end{figure}

\begin{figure}
	\centering
	\includegraphics{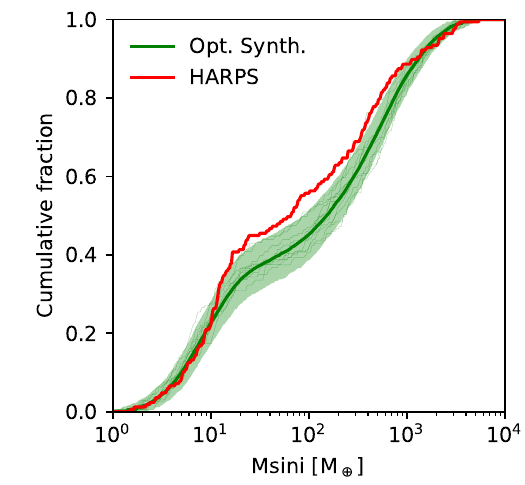}
	\caption{Cumulative distribution of the planet masses in the optimised synthetic population \texttt{NG192} (in green), the canonical 100~embryo population (in black), and the observed population (in red).}
	\label{fig:match-cumul-m}
\end{figure}

\begin{figure}
	\centering
	\includegraphics{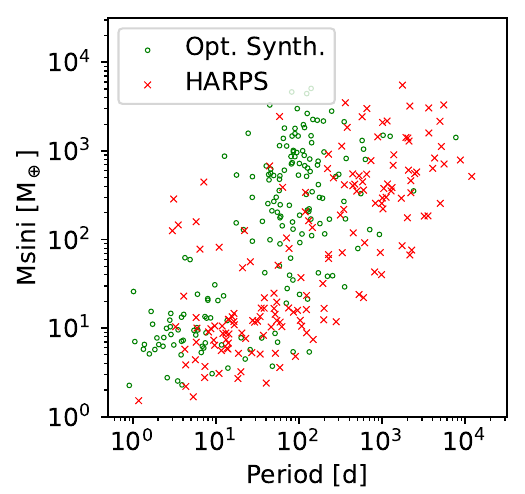}
	\caption{Mass-period diagram comparing one random mock observation of the optimised synthetic population \texttt{NG192} (in green) and the actual HARPS/Coralie planets (in red).}
	\label{fig:match-diag-pm}
\end{figure}

We highlight the effects of the planetesimals size and the overall efficiency of gas-driven migration in Appendix~\ref{sec:parameter-study}. Here, our goal is find a combination of parameters that best reproduces the total number of planets and their mass distribution (that is, not their location). The final optimised population for the HARPS/Coralie planets was obtained assuming the following parameters:
\begin{itemize}
	\item The planetesimals radius $\rplan$ has been set to \SI{2}{\kilo\meter} instead of \SI{300}{\meter} in the nominal population. This is mainly to reduce the efficiency of the model at forming giant planets.
	\item Gas-driven migration has been reduced to $7/8$ of its nominal rate, for both type~I and type~II migration.
	\item The maximum gas accretion rate is computed as the minimum of the Bondi rate and the radial flow of the gas disc, while in the canonical population it was the minimum of the Bondi rate and the local reservoir rate. This is to limit the gas accretion rate in the detached phase and better reproduce the masses of the giant planets.
\end{itemize}
The resulting population, which we refer to as \texttt{NG192}, contains \num{1000} systems and has been analysed in the same manner as the original one presented in the previous section. The histogram of the number of observed planets per system comparing this population with the HARPS/Coralie data is provided in Fig.~\ref{fig:match-mult}, the cumulative distribution of the planet masses in Fig.~\ref{fig:match-cumul-m}, and the mass-period diagram of one draw in Fig.~\ref{fig:match-diag-pm}.

The comparison of the number of planets in the new population provides are nearly matching those of the HARPS/Coralie systems, as there are $179_{-20}^{+22}$ planets in $120_{-12}^{+13}$ systems (as reminder HARPS/Coralie has 169 planets and candidates in 105 systems). Also, Fig.~\ref{fig:match-mult} shows that the distribution of system multiplicity is a good match with the combined HARPS/Coralie sample, although the synthetic populations shows too many systems with two planets. Unlike the nominal population, however, this population can produce one system with 6 observed planets (as indicated by the green region on the bottom right of Fig.~\ref{fig:match-mult}), albeit not in the majority of the samples.

The planetary mass function of the new population better reproduces the observed population overall. The remaining discrepancy lies with the most massive planets (more than 10 Jovian masses), which the synthetic population fails to reproduce. Thus, we note that it is difficult to find a prescription for the maximum gas accretion rate that is able to produce at the same time giants with an occurrence peak at about \SI{500}{\mearth} and the presence of very massive giant planets. Nevertheless, the KS distance at the \SI{95}{\percent} of the random observation is \num{1.35} compared to a limit value of \num{1.36}, which means that we are just below the rejection threshold for the hypothesis that the planets are drawn from the same sample.

However, the improvements in the reproduction of the planet mass function are not reflected in all the trends we investigated in the previous section. For instance, the planets are still closer-in as in the nominal population (Fig.~\ref{fig:match-diag-pm}). This is reflected in the 2D (mass-period) KS test, as we find that the distance of the observed sample is larger than any of the synthetic samples, meaning that the $p<0.001$ for the hypothesis that the observed sample is drawn for the synthetic population. We thus find that our model is unable to produce a synthetic population that reproduces at the same time the number of planets and their mass and period distributions. The issue here is that the reduction of the model's efficiency by increasing the size of the planetesimals also lengthen the formation time. The increased formation time make the planets more subject to planetary migration. Thus, the strongly coupled nature of the problem implies that it is generally not possible to ``force'' one outcome in one direction without also introducing many other consequences.

Also, the eccentricities are even lower than in the nominal population, as there are fewer massive planets (which are the source of excitation in our model). Conversely, the metallicity effect becomes stronger, as the decreased efficiency result in more selective formation around higher-metallicity stars.

Other possible improvements lie in a better understanding of the observed sample. This could be used to better inform the model about its properties and thus having initial conditions for the synthetic systems that better resembles the observations. Avenues in these directions include having a distribution of stellar masses, binary stars, and different elemental composition.

\section{Summary and conclusion}
\label{sec:conclusion}

In this work, we have compared the synthetic planet population obtained in \papertwo{} with the large HARPS/Coralie RV survey \citepalias{2011MayorArxiv}. We applied an observational bias derived from the survey to perform quantitative comparisons. The purpose of this endeavour was to statistically confront model predictions with observations and determine whether some physical processes that occur during planet formation are not yet well understood \citep{2009A&AMordasinib} and which aspects can in contrast be reproduced.

We find that the nominal synthetic population reproduces several of the main characteristics of the HARPS sample, namely, the two clusters of close-in sub-Neptunes and more distant giants in the mass-distance diagram; a bimodal mass function; an observed mean multiplicity of about 1.6; and several important trends in the stellar metallicity dependency, period ratio distribution, and eccentricity distribution. This could indicate that the model contains at least some of the physical mechanisms governing  planetary formation. This is a remarkable result considering that the model was not optimised beforehand to reproduce any particular survey. Instead, the model starts from 100 moon-mass embryos per disc, with all following an evolution governed by the physics included in the model. The narrower discrepancies are informative for the calibration of the various processes in the model and the identification of missing physics, and we discuss them in this section. To be able to observe these differences, in our work it is crucial that we can make a quantitative, and not only qualitative, comparison between theory and observation.

First, in terms of the absolute number of detectable planets, the nominal synthetic population has a greater number of planets by a factor of 1.7 compared to the observed sample. Thus, the nominal formation model is too efficient, albeit not by as much as found in \citet{2019ApJMulders} for the Kepler survey. This could be linked to model parameters controlling the efficiency of solid accretion or the initial conditions. The second discrepancy concerns the mass distribution of planets. We find that 1) the desert of planet masses between that of Neptune and Saturn is too deep by about \SI{60}{\percent}, 2) the mass of the peaks do not match exactly -- synthetic sub-Neptunes have too-low masses, while synthetic giants are too massive by a factor of about two in both cases -- and 3) the ratio between giants and sub-Neptunes is too large in the synthetic population~\citep[compare][NGPPS III]{2021AASchleckerA}. A third discrepancy relates to the location of the planets: The synthetic planets are overall too close to the star compared to the HARPS sample.

To better reproduce the number and mass distribution of planets in the synthetic population, we study in Appendix~\ref{sec:parameter-study} the effects of two parameters: the planetesimal size and the efficiency of gas-driven migration. In addition, we modify the gas accretion prescription to include the radial flow of gas through the disc, but it has little effect on the number or planet masses (Fig.~\ref{fig:psize-comp}).

Increasing the planetesimal size reduces the overall number of observed planets. In addition, this leads to an increase in the relative amount of sub-Neptunes compared to giants. Reducing the efficiency of gas-driven migration also changes the relative amount of sub-Neptunes and giants, with lower migration rates resulting in more giant planets and fewer sub-Neptunes. In our model, detectable sub-Neptunes and giants originate from the same region (beyond the ice line) and migrate to their final location. In this situation, the question of whether a planetary embryo becomes a sub-Neptune or a giant depends on the relative timescales of accretion versus migration \citep{2023EPJPEmsenhuber}. If migration is dominant (for instance with larger planetesimals, since that reduces the solid accretion rate), then sub-Neptunes are favoured, while the opposite leads to more giant planets as the study of the effects of migration efficiency has shown. This illustrates that modifying model parameters will affect not only a single outcome of the model but may also have consequences for other observable quantities in a way that might not be desirable for the goal of concurrently reproducing as many observational constraints as possible.

This outcome is caused by the many links and the feedback between the different physical processes that are included in the model (see the many arrows in Fig. 2 of \paperone{}). It implies that despite the many free parameters in a global model, it is difficult to force the model's outcome towards a desired direction just by varying parameters. This also shows that the depth of the planetary desert is not determined by runaway gas alone. The more efficient the migration is, the less amount of time planets can remain in this intermediate mass range without being taken to the inner edge of the disc. This finding has implications for how to analyse the HARPS/Coralie planets. The fact that the vast majority of the observable planets in our model come from a similar region also means that the whole population must be analysed together, as opposed to analysing sub-Neptunes and giant planets separately.

We find that larger planetesimals and a slight reduction of gas-driven planetary migration enable better reproduction of the masses of the HARPS/Coralie planets. However, the synthetic planets generally remain too close-in. Yet, migration cannot be reduced further, as was suggested by \citet{2018ApJIda} because, at least in our model, that would also lead to the formation of too many giant planets. This is a case where just varying the model parameters is insufficient to bring the syntheses into full agreement with the combined observational constraints, and it may suggest that planets originate from further away than they do in our model. However, that is problematic solely with planetesimal accretion, which suffers from long accretion timescales at large separations. Thus, it is an indication of missing physics in the modal and that a process such as pebble accretion \citep{2010AAOrmelKlahr} and/or structured discs \citep{2022AALau} might be at play for giant planet formation.

Regarding the constraints beyond the masses and orbital distances, we analysed the distribution of the period ratios of adjacent planets (Fig.~\ref{fig:comp-pratio}) and the eccentricity distribution (Fig.~\ref{fig:ecc-cumul}). Regarding the period ratios, the overall synthetic and observed distributions are in good agreement, with the exception of the detailed structure of the pile-up of period ratios past the location of the 2:1 MMR. While the overall number of planets in or near this pile-up is similar, the period ratios occupy a much narrower region (close to exactly 2.0) in the synthetic population than in the HARPS/Coralie planets (up to about 2.5).

Regarding the eccentricity distribution, the median eccentricity of the detectable synthetic planets is 0.07, compared to 0.15 in the HARPS/Coralie sample. In other words, the synthetic population contains a significant number of planets with circular or nearly circular orbits. Apart from this overall shift, the distributions are  similar.

Taken together, the period ratios and the eccentricities consistently show that the synthetic systems are too dynamically cold compared to HARPS/Coralie planets. This remains true even when the N-body run-time is extended to \SI{100}{\mega\year}  in the simulations (compared to \papertwo{}, where the interactions were followed up to only \SI{20}{\mega\year}). This indicates that we are missing an effect for the excitation of the planet orbits and that this effects occurs either early during the formation stage or has an external origin. The discrepancy could thus be caused either by a too strong eccentricity and inclination damping while the planets are embedded in the gas disc, as was found by \citet{2020AABitsch}. External excitation by stellar encounters could also be an explanation \citep{2011MNRASMalmberg}. There is an additional alternative explanation on the observational rather than the model side: It could also be that the eccentricities in the HARPS/Coralie sample are overestimated given the well-known bias of the RV method \citep{1971AJLucySweeney}.

Finally, the links to the host star properties can also be studied and contrasted. Concerning the metallicity of the systems harbouring observable planets, we find that both nominal and modified populations reproduce the overall  metallicity dependence of the planet mass quite well but differ again on a finer scale. Namely, in the model the correlation is weaker, though only two out of eight mass bins do not fit the observational 95\% confidence interval. The difference stems from sub-Neptune planets that have too-high median metallicities in the model.

The reduced planet growth efficiency when increasing the planetesimal size results in a higher fraction of systems coming from high-metallicity systems. This affects all masses, however, such that the median metallicity increases from \num{0.09} to \num{0.14} without significantly affecting the mass dependence. This means that our model favours overall high-metallicity systems for both sub-Neptunes and giant planets, and we find it difficult to form the close-packed systems with many sub-Neptunes at low metallicity (HD 40307; \citealp{2013AATuomi}, HD 136352; \citealp{2019AAUdry}). Again, this points to physical processes that are currently missing in the model, such as mechanisms affecting the formation of the solid building blocks   \citep{2021AAVoelkel,2022AAVoelkel} or the effect of envelope enrichment \citep[e.g.][]{2016AAVenturini,2024AAMolLous}.

Looking ahead, many large space- and ground-based instruments and surveys (e.g. NIRPS, GAIA, PLATO, ROMAN, ARIEL, etc.) will release observational datasets that are ideal for future statistical comparison with theoretical population-level predictions due to the homogeneous way the data are obtained \citep{exopag2023}. Theoretical planet formation models should use these surveys as an opportunity to go beyond the qualitative comparison of specific aspects and perform quantitative statistical comparison in a global way, that is, by comparing the same synthetic population in many aspects to all the different observational datasets. This will reveal the success and limitations of our current understanding of planet formation much more clearly and help answer the many fundamental questions still open \citep{2024RvMGMordasiniBurn}.

\begin{acknowledgements}
    We thank the anonymous reviewer, whose comments helped improve the quality of the manuscript.
    A.E. and C.M. acknowledge support from the Swiss National Science Foundation under grant 200021\_204847 ``PlanetsInTime''.
	A.E. and E.A. acknowledge the support from The University of Arizona.
	Parts of this work have been carried out within the framework of the NCCR PlanetS supported by the Swiss National Science Foundation under grants 51NF40\_182901 and 51NF40\_205606.
    R.B. acknowledges the financial support from DFG under Germany’s Excellence Strategy EXC 2181/1-390900948, Exploratory project EP 8.4 (the Heidelberg STRUCTURES Excellence Cluster).
	The results reported herein benefited from collaborations and/or information exchange within the program “Alien Earths” (supported by the National Aeronautics and Space Administration under agreement No. 80NSSC21K0593) for NASA’s Nexus for Exoplanet System Science (NExSS) research coordination network sponsored by NASA’s Science Mission Directorate.
	Calculations were performed on the Horus cluster at the University of Bern.
	The plots shown in this work were generated using matplotlib \citep{2007CSEHunter}.
\end{acknowledgements}

\bibliographystyle{aa}
\bibliography{manu,add}

\begin{appendix}

\section{Effect of the initial number of embryos}
\label{sec:nemb}

\papertwo{} presented multiple populations with different initial number of embryos per system: 1, 10, 20, 50, and 100. It was found that the giant planet population was relatively independent of the initial number of embryos per system as long as there are at least 10, except for a slight shift in the location of the giants. The 100-embryo population was taken as the nominal population, and this is the one that is mainly analysed here. However, obtaining such a population requires large computational resources, such that a parameter study is unfeasible. To perform such a parameter study with the available computational resources, we have to lower the number of embryos per system to 20.

\begin{figure}[h]
	\centering
	\includegraphics{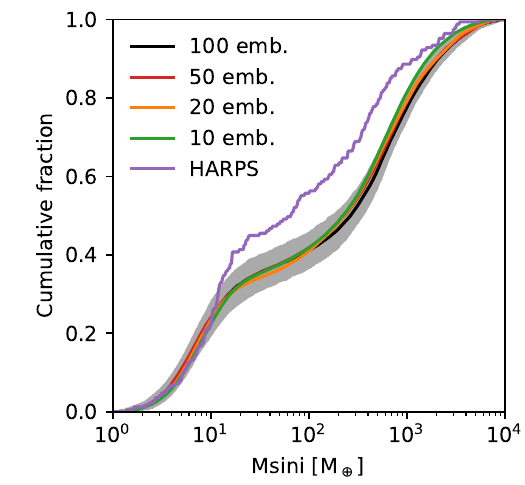}
	\caption{Cumulative distribution of the planet masses for biased synthetic populations with different initial number of embryos per system.}
	\label{fig:nemb}
\end{figure}

To check the effect of the initial number of embryos for comparison with HARPS, we consider the cumulative distribution of the observed planet masses, which is shown in Fig.~\ref{fig:nemb}. The comparison shows that the observed mass distribution is insensitive  to that parameter, except for the masses of the most massive planets, which are lower in the population with fewer embryos initially. Nevertheless, the difference is of the order of the confidence interval and lower than the difference with the observed planets.

Thus, we confirm the discussion in \papertwo{} that the initial number of embryos is significant only for the close-in terrestrial planets. As virtually none of these planets are observable by HARPS because of their too low mass, this model parameter has no significant effects on the results presented in this work. This also means that selecting a lower initial number of embryos for the populations that are used to find a better match with the HARPS sample should not cause any issue.

\FloatBarrier

\section{Parameter study}
\label{sec:parameter-study}

To find a population that is optimised for the HARPS/Coralie planets in absolute number and mass distribution, we perform a parameter study with several of the model parameters. Here we present the main results of that study and highlight how we selected the population shown in Sect.~\ref{sec:opt}.

\subsection{Planetesimal size}

The general efficiency of planet formation in the present model is controlled by the size of the planetesimals. For a given surface density, larger planetesimals would lead to a lower growth rate due to, first, the reduced damping of eccentricities and inclinations by the nebular gas and, second, a smaller enhancement of the capture cross-section by the planet envelope \citep{1988IcarusPodolak}.

\begin{figure}[h]
	\centering
	\includegraphics{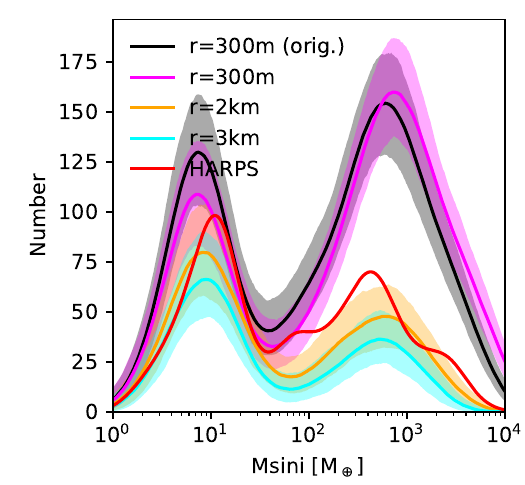}
	\caption{Comparison of the planet mass distribution between different biased synthetic populations and the observed population. The nominal population with 20 embryos per system (\texttt{NG74}, same as the orange line in Fig.~\ref{fig:nemb}) is shown with the black line; the others use the minimum of the Bondi rate and radial flow of the gas disc for the disc-limited gas accretion rate with different planetesimal size, as given in the legend. For each of the synthetic populations, we performed \num{1000} Monte Carlo synthetic observations. The bold line showing the median of these, and the lighter region shows the \SI{95}{\percent} confidence interval. The results are shown as a kernel density estimate, though rescaled to be absolute.}
	\label{fig:psize-comp}
\end{figure}

The results of this study are shown in Fig.~\ref{fig:psize-comp}, which compare the 20-embryos counterpart of the nominal population (called \texttt{NG74} in \papertwo; to keep this parameter the same across the comparison) to other populations where the maximum gas accretion rate criterion is given by the minimum of the Bondi rate and the radial flow of the gas and various planetesimals sizes. We see that the number of giant planets is much more affected by the change of planetesimal size than for the sub-Neptunes. For \SI{300}{\meter} planetesimals, giants are dominant, while for \SI{2}{\kilo\meter} and \SI{3}{\kilo\meter} the sub-Neptunes are. Increasing the planetesimal size thus solves both the issue of the number of planets and the imbalance between giants and sub-Neptunes. However, the best size for both items is not exactly the same: The population with \SI{2}{\kilo\meter} planetesimals shows an overall good match for the number of planets while having a small skew towards sub-Neptunes. For the latter point, smaller planetesimals would be beneficial.

Changing the planetesimal size has some additional consequences. For instance the location of the desert slightly shifts towards larger masses as the planetesimal size increases, while the location of the peak of the mass function for giant planets shifts towards lower masses. The latter is due to the longer formation time required with larger planetesimals, resulting in giant planets spending less time in the runaway gas accretion regime.

\subsection{Efficiency of migration}

\begin{figure}[h]
	\centering
	\includegraphics{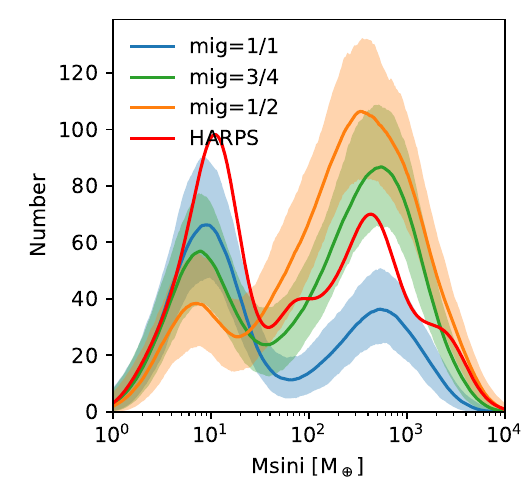}
	\caption{Same as Fig.~\ref{fig:psize-comp} but now comparing the effects of the overall efficiency of gas driven migration by multiplying the migration rates by uniform factors given in the legend. All the populations have a planetesimal size of \SI{3}{\kilo\meter}. The population shown in blue (labelled `1/1') is the same as that in cyan in Fig.~\ref{fig:psize-comp}.}
	\label{fig:mig-comp}
\end{figure}

As discussed in the Appendix of \papertwo{}, \citet{2021AASchleckerA} (NGPPS III), and in Sect.~\ref{sec:nominal-dist}, the synthetic planets are generally too close-in. Hence, we also study if a reduction of the gas-driven migration is able to better reproduce the location of to the observed planets. However, migration not only affects the planets' locations but also all their formation sequence. Therefore, we also study the effect of planetary migration on the mass function of the synthetic planets. The results of this comparison are shown in Fig.~\ref{fig:mig-comp}, where we show the mass function obtained by multiplying the migration rate by uniform factors.

Suppressing planet migration allows for more efficient giant planet formation, as already found in simulated planetary systems around very low-mass stars \citep{2022AASchlecker}. This skews the distribution of planet masses towards giants. In a sense, it is similar to reducing the planetesimal size, though with limited effects on the overall number of planets. This means that to avoid having too many giant planets, migration cannot be strongly reduced. Even a reduction by \SI{25}{\percent} (the `3/4' population) leads to too many giant planets. A modest reduction of the order of \SI{10}{\percent}, however, is beneficial to solving the discrepancy about the optimal planetesimals size for the overall number of planets and the ratio between sub-Neptunes and giants.

As for the planetesimal size, changing the efficiency of gas-driven migration affects the location of the planetary desert and the peak of the mass function of giant planets. There is a positive correlation between migration strength and the location of the gap and the peak. The location of the desert is best reproduced for a migration strength that is nominal or only slightly reduced, while that of the peak could be reproduced with weaker migration. Thus, the slight reduction by about \SI{10}{\percent} is also beneficial for reproducing the location of gaps and peaks in the planetary mass function.

\FloatBarrier

\section{Survey properties and list of planets}
\label{sec:plans}

The observed planets used for the comparison in this work are based on the results of the CORALIE and HARPS surveys until the end of 2015. Data from other instruments, such as HIRES/Keck, was not used. The original stellar sample was defined as volume-limited within \SI{50}{pc} as originally defined in \citet{2003MsngrMayor}. Newer measurements of stellar distances, for instance with GAIA parallaxes, were not used alter the sample. This is to maintain the consistency in the baseline duration of the observations across the stars.

\begin{figure}
	\centering
	\includegraphics{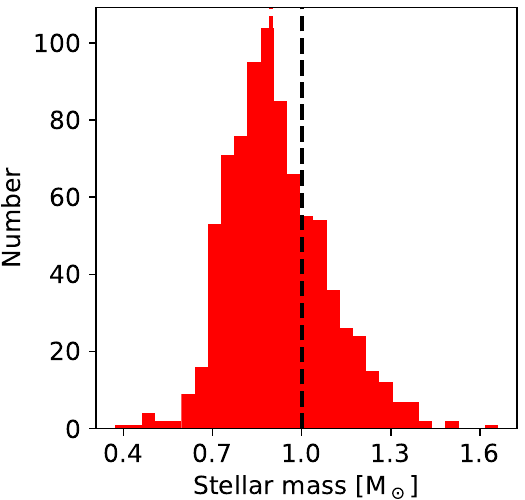}
	\caption{Histogram of the stellar masses in the HARPS/Coralie sample. The median value is indicated with the vertical dashed red line, while the value used for the synthetic population is indicated with the vertical dashed red line.}
	\label{fig:stellar-masses}
\end{figure}

The distribution of stellar masses in the sample is provided in Fig.~\ref{fig:stellar-masses}. The vast majority stars in the survey have masses between \num{0.7} and \SI{1.3}{\msun}. The distribution peaks around \SI{0.9}{\msun}, with a median value of \SI{0.90}{\msun} and a mean value of \SI{0.92}{\msun}.

Binaries, as indicated by a detectable companion within 6 arcsecond or a RV signal that indicate a companion of more than \SI{13}{\mj} were also excluded. While this criterion remove most binaries, it does not exclude wide binary systems. Among the detected planets, 24 (\SI{14}{\percent} of the total) of them in 20 systems (\SI{19}{\percent}) are in known binary systems.

With the exclusion of the additional four years of data, the analysis was performed as in \citet{2011MayorArxiv}. The analysis was performed consistently across the entire sample and the results are reported in Table~\ref{tab:plans}. For long period planets with incomplete coverage, the eccentricity is forced to zero if it cannot be reliably determined, and this is then used to determine the other parameters.

The data used here may differ from that presented in the discovery publications. For comparison, we provide the values from the literature. Values that differ more than \SI{10}{\percent} are underlined. Usually, differences in the periods are minimal, while differences in masses can be larger. However, as we are interested in the statistical comparison in this work, this should not be a major concern here.

In addition, the tables also provides the metallicity of the host star. Except for certain stars, the values were taken from SWEET-Cat. In case the entry does not provide a reference for the stellar metallicity, the value was taken from the same reference as for the planet.

\onecolumn

\begin{landscape}
\begin{longtable}{lllllllrlll}
\caption{List of planets from the HARPS/Coralie surveys}
\label{tab:plans} \\
\hline
Designation & Period & Ecc.  & $M\sin{i}$    & Period' & Ecc.' & $M\sin{i}'$   & [Fe/H] & Ref. discovery & Ref. planet & Ref. [Fe/H] \\
            & [d]    &       & [\si{\mearth}] & [d]     &       & [\si{\mearth}] & & & &\\
\hline
\endhead
HD142b & \num{350.3} & \num{0.25} & \num{415} & \num{350.3} & \num{0.25} & \num{415} & \num{0.09} & \citet{2002ApJTinney} & \citet{2002ApJTinney} & \citet{2008AASousa} \\
HD142c & \num{6005} & \num{0.21} & \num{1684} & \num{6005} & \num{0.21} & \num{1684} & \num{0.09} & \citet{2012ApJWittenmyer} & \citet{2012ApJWittenmyer} & \citet{2008AASousa} \\
HD1461b & \num{5.77} & \ldots & \num{6.94} & \num{5.77} & \num{<0.17} & \num{6.44} & \num{0.19} & \citet{2010ApJRivera} & \citet{2016AADiaz} & \citet{2008AASousa} \\
HD1461c & \num{13.50} & \ldots & \num{5.92} & \num{13.51} & \num{<0.31} & \num{5.59} & \num{0.19} & \citet{2016AADiaz} & \citet{2016AADiaz} & \citet{2008AASousa} \\
HD4113b & \num{526.62} & \num{0.90} & \num{523.9} & \num{526.62} & \num{0.90} & \num{495.79} & \num{0.20} & \citet{2008AATamuz} & \citet{2008AATamuz} & \citet{2008AATamuz} \\
HD4208b & \num{828} & \num{0.05} & \num{256.6} & \num{833} & \underline{\num{0.04}} & \num{257} & \num{-0.28} & \citet{2002ApJVogt} & \citet{2018AABarbato} & \citet{2008AASousa} \\
HD4308b & \num{15.62} & \num{0.08} & \num{13.45} & \num{15.56} & \underline{\num{0.00}} & \num{14.05} & \num{-0.34} & \citet{2006AAUdry} & \citet{2006AAUdry} & \citet{2008AASousa} \\
HD6434b & \num{21.99} & \num{0.17} & \num{126.2} & \num{22.00} & \num{0.17} & \num{123.95} & \num{-0.53} & \citet{2004AAMayor} & \citet{2004AAMayor} & \citet{2008AASousa} \\
HD7199b & \num{615} & \num{0.19} & \num{92} & \num{615} & \num{0.19} & \num{92} & \num{0.28} & \citet{2011AADumusque} & \citet{2011AADumusque} & \citet{2011AADumusque} \\
HD7449b & \num{1275} & \num{0.82} & \num{353} & \num{1275} & \num{0.82} & \num{353} & \num{-0.11} & \citet{2011AADumusque} & \citet{2011AADumusque} & \citet{2011AADumusque} \\
HD7449c & \num{4046} & \num{0.53} & \num{636} & \num{4046} & \num{0.53} & \num{636} & \num{-0.11} & \citet{2011AADumusque} & \citet{2011AADumusque} & \citet{2011AADumusque} \\
HD10180b & \num{1.17} & \ldots & \num{1.52} & \num{1.18} & \ldots & \underline{\num{1.35}} & \num{0.08} & \citet{2011AALovis} & \citet{2011AALovis} & \citet{2008AASousa} \\
HD10180c & \num{5.75} & \num{0.07} & \num{13.2} & \num{5.76} & \underline{\num{0.05}} & \num{13.1} & \num{0.08} & \citet{2011AALovis} & \citet{2011AALovis} & \citet{2008AASousa} \\
HD10180d & \num{16.36} & \num{0.16} & \num{11.9} & \num{16.36} & \underline{\num{0.09}} & \num{11.8} & \num{0.08} & \citet{2011AALovis} & \citet{2011AALovis} & \citet{2008AASousa} \\
HD10180e & \num{49.75} & \num{0.06} & \num{24.8} & \num{49.75} & \underline{\num{0.03}} & \num{25.1} & \num{0.08} & \citet{2011AALovis} & \citet{2011AALovis} & \citet{2008AASousa} \\
HD10180f & \num{122.7} & \num{0.13} & \num{23.4} & \num{122.76} & \num{0.14} & \num{23.9} & \num{0.08} & \citet{2011AALovis} & \citet{2011AALovis} & \citet{2008AASousa} \\
HD10180g & \num{595} & \ldots & \num{22.1} & \num{601} & \num{0.19} & \num{21.4} & \num{0.08} & \citet{2011AALovis} & \citet{2011AALovis} & \citet{2008AASousa} \\
HD10180h & \num{2149} & \num{0.15} & \num{67} & \num{2222} & \underline{\num{0.08}} & \num{64.4} & \num{0.08} & \citet{2011AALovis} & \citet{2011AALovis} & \citet{2008AASousa} \\
HD10647b & \num{1003} & \num{0.15} & \num{294} & \num{989} & \num{0.15} & \num{298} & \num{0.00} & \citet{2006ApJButler} & \citet{2013AAMarmier} & \citet{2008AASousa} \\
HD11964Ac & \num{37.94} & \num{0.21} & \num{16.8} & \num{37.94} & \num{0.21} & \num{16.8} & \num{0.10} & \citet{2009ApJWright} & \citet{2009ApJWright} & \citet{2018AASousa} \\
HD11964Ab & \num{2010} & \num{0.07} & \num{176} & \num{2010} & \num{0.07} & \num{176} & \num{0.10} & \citet{2006ApJButler} & \citet{2009ApJWright} & \citet{2018AASousa} \\
HD13808b & \num{14.18} & \ldots & \num{12.6} & \num{14.18} & \num{0.07} & \underline{\num{11.2}} & \num{-0.21} & \citet{2021MNRASAhrer} & \citet{2021MNRASAhrer} & \citet{2013AATsantaki} \\
HD13808c & \num{53.74} & \ldots & \num{12.2} & \num{53.75} & \num{0.16} & \underline{\num{9.96}} & \num{-0.21} & \citet{2021MNRASAhrer} & \citet{2021MNRASAhrer} & \citet{2013AATsantaki} \\
HD16141b & \num{75.52} & \num{0.25} & \num{79.3} & \num{75.46} & \underline{\num{0.20}} & \num{79.5} & \num{0.16} & \citet{2000ApJMarcy} & \citet{2022ApJSRosenthal} & \citet{2008AASousa} \\
HD16417b & \num{16.43} & \num{0.22} & \num{14.7} & \num{17.24} & \num{0.20} & \underline{\num{22.1}} & \num{0.13} & \citet{2009ApJOToole} & \citet{2009ApJOToole} & \citet{2008AASousa} \\
HD19994b & \num{466.2} & \num{0.26} & \num{421.7} & \underline{\num{535.7}} & \underline{\num{0.30}} & \underline{\num{534}} & \num{0.24} & \citet{2004AAMayor} & \citet{2004AAMayor} & \citet{2008AASousa} \\
HD20003b & \num{11.84} & \num{0.34} & \num{10.57} & \num{11.84} & \num{0.36} & \num{11.66} & \num{0.04} & \citet{2019AAUdry} & \citet{2019AAUdry} & \citet{2008AASousa} \\
HD20003c & \num{33.85} & \ldots & \num{13.51} & \num{33.92} & \num{0.10} & \num{14.44} & \num{0.04} & \citet{2019AAUdry} & \citet{2019AAUdry} & \citet{2008AASousa} \\
HD20781b & \num{5.31} & \num{0.02} & \num{1.68} & \num{5.31} & \underline{\num{0.10}} & \underline{\num{1.93}} & \num{-0.11} & \citet{2019AAUdry} & \citet{2019AAUdry} & \citet{2008AASousa} \\
HD20781c & \num{13.89} & \num{0.03} & \num{5.82} & \num{13.89} & \underline{\num{0.09}} & \num{5.33} & \num{-0.11} & \citet{2019AAUdry} & \citet{2019AAUdry} & \citet{2008AASousa} \\
HD20781d & \num{29.15} & \num{0.07} & \num{11.68} & \num{29.16} & \underline{\num{0.11}} & \num{10.61} & \num{-0.11} & \citet{2019AAUdry} & \citet{2019AAUdry} & \citet{2008AASousa} \\
HD20781e & \num{85.39} & \num{0.10} & \num{15.15} & \num{85.51} & \underline{\num{0.06}} & \num{14.03} & \num{-0.11} & \citet{2019AAUdry} & \citet{2019AAUdry} & \citet{2008AASousa} \\
HD20782b & \num{596.2} & \num{0.93} & \num{361} & \num{597.06} & \num{0.95} & \underline{\num{473}} & \num{-0.06} & \citet{2006MNRASJones} & \citet{2019AAUdry} & \citet{2008AASousa} \\
HD20794b & \num{18.31} & \ldots & \num{2.7} & \num{18.31} & \ldots & \num{2.7} & \num{-0.40} & \citet{2011AAPepe} & \citet{2011AAPepe} & \citet{2008AASousa} \\
HD20794c & \num{40.11} & \ldots & \num{2.4} & \num{40.11} & \ldots & \num{2.4} & \num{-0.40} & \citet{2011AAPepe} & \citet{2011AAPepe} & \citet{2008AASousa} \\
HD20794d & \num{90.30} & \ldots & \num{4.8} & \num{90.30} & \ldots & \num{4.8} & \num{-0.40} & \citet{2011AAPepe} & \citet{2011AAPepe} & \citet{2008AASousa} \\
HD21693b & \num{22.67} & \num{0.02} & \num{7.66} & \num{22.68} & \underline{\num{0.12}} & \num{8.23} & \num{0.00} & \citet{2019AAUdry} & \citet{2019AAUdry} & \citet{2008AASousa} \\
HD21693c & \num{53.83} & \num{0.12} & \num{19.68} & \num{53.74} & \underline{\num{0.07}} & \underline{\num{17.37}} & \num{0.00} & \citet{2019AAUdry} & \citet{2019AAUdry} & \citet{2008AASousa} \\
HD23079b & \num{730.6} & \num{0.10} & \num{776.6} & \num{730.6} & \num{0.10} & \num{778.7} & \num{-0.12} & \citet{2002ApJTinney} & \citet{2006ApJButler} & \citet{2008AASousa} \\
HD27631b & \num{2208} & \num{0.12} & \num{460.8} & \num{2198.14} & \underline{\num{0.14}} & \num{474.8} & \num{-0.11} & \citet{2013AAMarmier} & \citet{2018AABarbato} & \citet{2013AASantos} \\
HD28185b & \num{379} & \num{0.05} & \num{1842.5} & \num{383} & \underline{\num{0.07}} & \num{1812} & \num{0.21} & \citet{2001AASantosB} & \citet{2001AASantosB} & \citet{2008AASousa} \\
HD30562b & \num{1157} & \num{0.75} & \num{423.5} & \num{1159} & \num{0.77} & \num{436.36} & \num{0.28} & \citet{2009ApJFischer} & \citet{2013AAMarmier} & \citet{2018AASousa} \\
HD31527b & \num{16.55} & \num{0.05} & \num{10.91} & \num{16.55} & \underline{\num{0.10}} & \num{10.47} & \num{-0.17} & \citet{2019AAUdry} & \citet{2019AAUdry} & \citet{2008AASousa} \\
HD31527c & \num{51.25} & \num{0.01} & \num{14.37} & \num{51.21} & \underline{\num{0.04}} & \num{14.16} & \num{-0.17} & \citet{2019AAUdry} & \citet{2019AAUdry} & \citet{2008AASousa} \\
HD31527d & \num{279.02} & \num{0.07} & \num{11.20} & \num{271.67} & \underline{\num{0.24}} & \num{11.82} & \num{-0.17} & \citet{2019AAUdry} & \citet{2019AAUdry} & \citet{2008AASousa} \\
HD38858b & \num{197.6} & \ldots & \num{12.43} & \num{197.6} & \ldots & \num{12.43} & \num{-0.22} & \citet{2011MayorArxiv} & \citet{2011MayorArxiv} & \citet{2008AASousa} \\
HD39091b & \num{2151} & \num{0.64} & \num{3206.3} & \num{2151} & \num{0.64} & \num{3264.0} & \num{0.09} & \citet{2002MNRASJones} & \citet{2006ApJButler} & \citet{2008AASousa} \\
HD39194b & \num{5.63} & \ldots & \num{4.4} & \num{5.64} & \num{<0.11} & \num{4.0} & \num{-0.61} & \citet{2021AAUnger} & \citet{2021AAUnger} & \citet{2008AASousa} \\
HD39194c & \num{14.03} & \ldots & \num{6.8} & \num{14.03} & \num{<0.08} & \num{6.3} & \num{-0.61} & \citet{2021AAUnger} & \citet{2021AAUnger} & \citet{2008AASousa} \\
HD39194d & \num{33.89} & \ldots & \num{5.1} & \num{33.91} & \num{<0.17} & \underline{\num{4.0}} & \num{-0.61} & \citet{2021AAUnger} & \citet{2021AAUnger} & \citet{2008AASousa} \\
HD40307b & \num{4.31} & \num{0.12} & \num{3.85} & \num{4.31} & \underline{\num{<0.16}} & \num{3.81} & \num{-0.36} & \citet{2009AAMayorA} & \citet{2016AADiaz} & \citet{2013AATsantaki} \\
HD40307c & \num{9.61} & \num{0.06} & \num{6.42} & \num{9.62} & \underline{\num{<0.10}} & \num{6.43} & \num{-0.36} & \citet{2009AAMayorA} & \citet{2016AADiaz} & \citet{2013AATsantaki} \\
HD40307d & \num{20.43} & \num{0.05} & \num{8.82} & \num{20.42} & \underline{\num{<0.12}} & \num{8.74} & \num{-0.36} & \citet{2009AAMayorA} & \citet{2016AADiaz} & \citet{2013AATsantaki} \\
HD40307f & \num{51.76} & \num{0.02} & \num{5.21} & \num{51.56} & \underline{\num{<0.33}} & \underline{\num{3.63}} & \num{-0.36} & \citet{2013AATuomi} & \citet{2016AADiaz} & \citet{2013AATsantaki} \\
HD45184b & \num{5.88} & \num{0.09} & \num{10.02} & \num{5.88} & \underline{\num{0.07}} & \underline{\num{12.19}} & \num{0.04} & \citet{2019AAUdry} & \citet{2019AAUdry} & \citet{2008AASousa} \\
HD45184c & \num{13.12} & \num{0.10} & \num{8.81} & \num{13.14} & \underline{\num{0.07}} & \num{8.81} & \num{0.04} & \citet{2019AAUdry} & \citet{2019AAUdry} & \citet{2008AASousa} \\
HD45364b & \num{226} & \num{0.18} & \num{61.1} & \num{226.93} & \num{0.17} & \num{59.50} & \num{-0.17} & \citet{2009AACorreia} & \citet{2009AACorreia} & \citet{2008AASousa} \\
HD45364c & \num{344.3} & \num{0.12} & \num{219} & \num{342.85} & \underline{\num{0.10}} & \num{209.09} & \num{-0.17} & \citet{2009AACorreia} & \citet{2009AACorreia} & \citet{2008AASousa} \\
HD47186b & \num{4.08} & \num{0.04} & \num{23} & \num{4.08} & \num{0.04} & \num{22.78} & \num{0.23} & \citet{2009AABouchy} & \citet{2009AABouchy} & \citet{2008AASousa} \\
HD47186c & \num{3552} & \num{0.28} & \num{185} & \underline{\num{1354}} & \underline{\num{0.25}} & \underline{\num{111.42}} & \num{0.23} & \citet{2009AABouchy} & \citet{2009AABouchy} & \citet{2008AASousa} \\
HD50499b & \num{2457.87} & \num{0.25} & \num{554.5} & \num{2447} & \num{0.27} & \num{520.0} & \num{0.34} & \citet{2005ApJVogt} & \citet{2018AABarbato} & \citet{2018AASousa} \\
HD51608b & \num{14.07} & \num{0.09} & \num{12.48} & \num{14.07} & \num{0.09} & \num{12.77} & \num{-0.07} & \citet{2019AAUdry} & \citet{2019AAUdry} & \citet{2008AASousa} \\
HD51608c & \num{95.63} & \num{0.26} & \num{15.90} & \num{95.94} & \underline{\num{0.14}} & \underline{\num{14.31}} & \num{-0.07} & \citet{2019AAUdry} & \citet{2019AAUdry} & \citet{2008AASousa} \\
HD52265b & \num{119.29} & \num{0.32} & \num{340.5} & \num{119.26} & \underline{\num{0.21}} & \num{352.2} & \num{0.21} & \citet{2000ApJButler} & \citet{2022ApJSRosenthal} & \citet{2008AASousa} \\
HD65216b & \num{579} & \num{0.26} & \num{449} & \num{579} & \num{0.26} & \underline{\num{385}} & \num{-0.17} & \citet{2004AAMayor} & \citet{2004AAMayor} & \citet{2008AASousa} \\
HD65216c & \num{5542} & \num{0.15} & \num{712} & \num{5370} & \underline{\num{0.27}} & \num{645} & \num{-0.17} & \citet{2019MNRASWittenmyer} & \citet{2019MNRASWittenmyer} & \citet{2008AASousa} \\
HD69830b & \num{8.66} & \num{0.10} & \num{9.86} & \num{8.67} & \num{0.10} & \num{10.2} & \num{-0.06} & \citet{2006NatureLovis} & \citet{2006NatureLovis} & \citet{2008AASousa} \\
HD69830c & \num{31.58} & \num{0.10} & \num{11.98} & \num{31.56} & \underline{\num{0.13}} & \num{11.8} & \num{-0.06} & \citet{2006NatureLovis} & \citet{2006NatureLovis} & \citet{2008AASousa} \\
HD69830d & \num{199.56} & \num{0.16} & \num{16.73} & \num{197} & \underline{\num{0.07}} & \num{18.1} & \num{-0.06} & \citet{2006NatureLovis} & \citet{2006NatureLovis} & \citet{2008AASousa} \\
HD70642b & \num{2068} & \num{0.03} & \num{606.9} & \num{2124} & \underline{\num{0.18}} & \num{633.4} & \num{0.18} & \citet{2003ApJCarter} & \citet{2018AABarbato} & \citet{2008AASousa} \\
HD75289b & \num{3.50} & \num{0.03} & \num{146.2} & \num{3.51} & \underline{\num{0.02}} & \num{133} & \num{0.38} & \citet{2000AAUdry} & \citet{2000AAUdry} & \citet{2008AASousa} \\
HD77338b & \num{5.75} & \num{0.} & \num{15.88} & \num{5.75} & \num{0.} & \num{15.9} & \num{0.28} & \citet{2013ApJJenkins} & \citet{2013ApJJenkins} & \citet{2008AASousa} \\
HD82943c & \num{219.5} & \num{0.35} & \num{632.2} & \num{219.4} & \num{0.38} & \num{588} & \num{0.26} & \citet{2004AAMayor} & \citet{2004AAMayor} & \citet{2008AASousa} \\
HD82943b & \num{441.2} & \num{0.21} & \num{551} & \num{435.1} & \underline{\num{0.18}} & \num{585} & \num{0.26} & \citet{2004AAMayor} & \citet{2004AAMayor} & \citet{2008AASousa} \\
HD83443b & \num{2.98} & \num{0.01} & \num{125.8} & \num{2.99} & \underline{\num{0.07}} & \num{130} & \num{0.34} & \citet{2002ApJButler} & \citet{2022ApJSRosenthal} & \citet{2008AASousa} \\
HD85390b & \num{809.4} & \num{0.44} & \num{43.1} & \num{788} & \num{0.41} & \num{42} & \num{-0.09} & \citet{2011AAMordasini} & \citet{2011AAMordasini} & \citet{2013AATsantaki} \\
HD85512b & \num{58.43} & \num{0.11} & \num{3.6} & \num{58.43} & \num{0.11} & \num{3.6} & \num{-0.26} & \citet{2011AAPepe} & \citet{2011AAPepe} & \citet{2013AATsantaki} \\
HD86226b & \num{1695} & \num{0.15} & \num{292.38} & \num{1695} & \num{0.15} & \num{292.39} & \num{0.02} & \citet{2009ApJMinniti} & \citet{2013AAMarmier} & \citet{2013AASantos} \\
HD90156b & \num{49.88} & \num{0.46} & \num{16.7} & \num{49.77} & \underline{\num{0.31}} & \num{18.0} & \num{-0.24} & \citet{2011AAMordasini} & \citet{2011AAMordasini} & \citet{2008AASousa} \\
HD92788b & \num{325.81} & \num{0.33} & \num{1132.7} & \num{325.71} & \num{0.36} & \num{1119} & \num{0.27} & \citet{2001ApJFischer} & \citet{2022ApJSRosenthal} & \citet{2008AASousa} \\
HD93083b & \num{144.2} & \num{0.07} & \num{136.5} & \num{143.58} & \underline{\num{0.14}} & \underline{\num{117.59}} & \num{0.08} & \citet{2005AALovis} & \citet{2005AALovis} & \citet{2013AATsantaki} \\
HD93385b & \num{7.34} & \num{0.12} & \num{3.7} & \num{7.34} & \underline{\num{<0.16}} & \underline{\num{4.2}} & \num{0.02} & \citet{2021AAUnger} & \citet{2021AAUnger} & \citet{2008AASousa} \\
HD93385c & \num{13.18} & \num{0.06} & \num{7.4} & \num{13.18} & \underline{\num{<0.10}} & \num{7.1} & \num{0.02} & \citet{2021AAUnger} & \citet{2021AAUnger} & \citet{2008AASousa} \\
HD93385d & \num{46} & \num{0.15} & \num{10.3} & \num{45.85} & \underline{\num{0.09}} & \underline{\num{8.7}} & \num{0.02} & \citet{2021AAUnger} & \citet{2021AAUnger} & \citet{2008AASousa} \\
HD96700b & \num{8.12} & \ldots & \num{9.08} & \num{8.12} & \num{<0.08} & \num{8.9} & \num{-0.18} & \citet{2021AAUnger} & \citet{2021AAUnger} & \citet{2008AASousa} \\
HD96700c & \num{19.89} & \ldots & \num{3.22} & \num{19.88} & \num{<0.14} & \num{3.5} & \num{-0.18} & \citet{2021AAUnger} & \citet{2021AAUnger} & \citet{2008AASousa} \\
HD96700d & \num{103.22} & \ldots & \num{12.25} & \num{103.5} & \num{<0.27} & \num{12.7} & \num{-0.18} & \citet{2021AAUnger} & \citet{2021AAUnger} & \citet{2008AASousa} \\
HD98649b & \num{4951} & \num{0.85} & \num{2161} & \underline{\num{6022}} & \num{0.86} & \num{2158} & \num{-0.03} & \citet{2013AAMarmier} & \citet{2019AARickman} & \citet{2013AASantos} \\
HD101930b & \num{70.49} & \num{0.07} & \num{104.2} & \num{70.46} & \underline{\num{0.11}} & \num{95.34} & \num{0.16} & \citet{2005AALovis} & \citet{2005AALovis} & \citet{2013AATsantaki} \\
HD102117b & \num{20.82} & \num{0.07} & \num{48} & \num{20.67} & \ldots & \num{44} & \num{0.28} & \citet{2005AALovis} & \citet{2005AALovis} & \citet{2008AASousa} \\
HD102365b & \num{122.1} & \num{0.34} & \num{16.2} & \num{122.1} & \num{0.34} & \num{16.0} & \num{-0.29} & \citet{2011ApJTinneyA} & \citet{2011ApJTinneyA} & \citet{2008AASousa} \\
HD104067b & \num{55.83} & \num{0.06} & \num{51.4} & \num{55.81} & \ldots & \underline{\num{59.0}} & \num{-0.04} & \citet{2011AASegransan} & \citet{2011AASegransan} & \citet{2013AATsantaki} \\
HD106515Ab & \num{3630} & \num{0.57} & \num{3054.1} & \num{3630} & \num{0.57} & \num{3054} & \num{0.03} & \citet{2013AAMarmier} & \citet{2013AAMarmier} & \citet{2013AASantos} \\
HD108147b & \num{10.89} & \num{0.52} & \num{82} & \num{10.90} & \num{0.50} & \underline{\num{127}} & \num{0.18} & \citet{2002AAPepe} & \citet{2002AAPepe} & \citet{2008AASousa} \\
HD111232b & \num{1143} & \num{0.20} & \num{2174.5} & \num{1143} & \num{0.20} & \num{2161} & \num{-0.43} & \citet{2004AAMayor} & \citet{2004AAMayor} & \citet{2008AASousa} \\
HD114386c & \num{447} & \num{0.12} & \num{119} & \ldots & \ldots & \ldots & \num{-0.09} & \citet{2011MayorArxiv} & \citet{2011MayorArxiv} & \citet{2013AATsantaki} \\
HD114386b & \num{1043} & \num{0.06} & \num{377} & \underline{\num{937.7}} & \num{0.06} & \num{394.09} & \num{-0.09} & \citet{2004AAMayor} & \citet{2004AAMayor} & \citet{2013AATsantaki} \\
HD114729b & \num{1114} & \num{0.16} & \num{300.3} & \num{1121.8} & \underline{\num{0.08}} & \underline{\num{262}} & \num{-0.28} & \citet{2003ApJButler} & \citet{2018AABarbato} & \citet{2008AASousa} \\
HD114783b & \num{493.7} & \num{0.14} & \num{351.2} & \num{492.56} & \num{0.13} & \num{328} & \num{0.03} & \citet{2002ApJVogt} & \citet{2022ApJSRosenthal} & \citet{2008AASousa} \\
HD115617b & \num{4.21} & \num{0.15} & \num{5.79} & \num{4.21} & \underline{\num{0.10}} & \underline{\num{5.12}} & \num{-0.01} & \citet{2010ApJVogt} & \citet{2022ApJSRosenthal} & \citet{2018AASousa} \\
HD115617c & \num{38.10} & \num{0.13} & \num{22.05} & \num{38.10} & \underline{\num{0.07}} & \underline{\num{16.1}} & \num{-0.01} & \citet{2010ApJVogt} & \citet{2022ApJSRosenthal} & \citet{2018AASousa} \\
HD115617d & \num{123.01} & \num{0.33} & \num{8.86} & \num{123.01} & \num{0.35} & \underline{\num{22.9}} & \num{-0.01} & \citet{2010ApJVogt} & \citet{2010ApJVogt} & \citet{2018AASousa} \\
HD117207b & \num{2597} & \num{0.14} & \num{578.1} & \num{2621.8} & \underline{\num{0.16}} & \num{612} & \num{0.22} & \citet{2005ApJMarcy} & \citet{2018AABarbato} & \citet{2008AASousa} \\
HD117618b & \num{25.82} & \num{0.41} & \num{56.1} & \num{25.80} & \num{0.45} & \num{55.3} & \num{0.03} & \citet{2005ApJTinney} & \citet{2019MNRASWittenmyer} & \citet{2008AASousa} \\
HD121504b & \num{63.33} & \num{0.02} & \num{388.5} & \num{63.33} & \underline{\num{0.03}} & \num{388} & \num{0.14} & \citet{2004AAMayor} & \citet{2004AAMayor} & \citet{2008AASousa} \\
HD126525b & \num{948.12} & \num{0.13} & \num{71.33} & \num{960.4} & \underline{\num{0.04}} & \num{75.3} & \num{-0.10} & \citet{2018AABarbato} & \citet{2018AABarbato} & \citet{2008AASousa} \\
HD134060b & \num{3.26} & \num{0.45} & \num{10.38} & \num{3.27} & \num{0.45} & \num{10.10} & \num{0.14} & \citet{2019AAUdry} & \citet{2019AAUdry} & \citet{2008AASousa} \\
HD134060c & \ldots & \ldots & \ldots & \num{1292} & \num{0.11} & \num{29.29} & \num{0.14} & \citet{2019AAUdry} & \citet{2019AAUdry} & \citet{2008AASousa} \\
HD134606e & \num{4.32} & \num{0.12} & \num{2.2} & \num{4.32} & \underline{\num{0.20}} & \num{2.34} & \num{0.27} & \citet{2024AJLi} & \citet{2024AJLi} & \citet{2008AASousa} \\
HD134606b & \num{12.09} & \num{0.06} & \num{8.7} & \num{12.09} & \underline{\num{0.09}} & \num{9.09} & \num{0.27} & \citet{2024AJLi} & \citet{2024AJLi} & \citet{2008AASousa} \\
HD134606f & \num{26.92} & \num{0.31} & \num{5.3} & \num{26.92} & \underline{\num{0.08}} & \num{5.63} & \num{0.27} & \citet{2024AJLi} & \citet{2024AJLi} & \citet{2008AASousa} \\
HD134606c & \num{58.97} & \num{0.07} & \num{11.7} & \num{58.88} & \underline{\num{0.06}} & \num{11.31} & \num{0.27} & \citet{2024AJLi} & \citet{2024AJLi} & \citet{2008AASousa} \\
HD134606g & \num{147.5} & \num{0.17} & \num{7.45} & \num{147.5} & \num{0.17} & \num{7.45} & \num{0.27} & \citet{2011MayorArxiv} & \citet{2011MayorArxiv} & \citet{2008AASousa} \\
HD134606d & \num{981.} & \num{0.08} & \num{40.19} & \num{966.5} & \underline{\num{0.09}} & \underline{\num{44.8}} & \num{0.27} & \citet{2024AJLi} & \citet{2024AJLi} & \citet{2008AASousa} \\
HD134987b & \num{258.18} & \num{0.23} & \num{496.9} & \num{258.25} & \num{0.23} & \num{516.2} & \num{0.25} & \citet{2000ApJVogt} & \citet{2022ApJSRosenthal} & \citet{2008AASousa} \\
HD134987c & \num{5000} & \num{0.11} & \num{255.8} & \underline{\num{5960}} & \underline{\num{0.15}} & \underline{\num{297}} & \num{0.25} & \citet{2010MNRASJones} & \citet{2022ApJSRosenthal} & \citet{2008AASousa} \\
HD136352b & \num{11.58} & \num{0.12} & \num{5.59} & \num{11.58} & \underline{\num{0.14}} & \underline{\num{4.81}} & \num{-0.34} & \citet{2019AAUdry} & \citet{2019AAUdry} & \citet{2008AASousa} \\
HD136352c & \num{27.56} & \num{0.07} & \num{12.22} & \num{27.58} & \underline{\num{0.04}} & \underline{\num{10.80}} & \num{-0.34} & \citet{2019AAUdry} & \citet{2019AAUdry} & \citet{2008AASousa} \\
HD136352d & \num{107.16} & \ldots & \num{10.55} & \num{107.60} & \num{0.09} & \underline{\num{8.58}} & \num{-0.34} & \citet{2019AAUdry} & \citet{2019AAUdry} & \citet{2008AASousa} \\
HD137388b & \num{330} & \num{0.36} & \num{71} & \num{330} & \num{0.36} & \num{70.9} & \num{0.18} & \citet{2011AADumusque} & \citet{2011AADumusque} & \citet{2011AADumusque} \\
HD141937b & \num{653.21} & \num{0.40} & \num{3011.5} & \num{653.22} & \num{0.41} & \num{3080} & \num{0.13} & \citet{2002AAUdry} & \citet{2002AAUdry} & \citet{2008AASousa} \\
HD142022Ab & \num{1928} & \num{0.52} & \num{1419.9} & \num{1928} & \num{0.53} & \underline{\num{1600}} & \num{0.19} & \citet{2006AAEggenberger} & \citet{2006AAEggenberger} & \citet{2008AASousa} \\
HD147018b & \num{44.23} & \num{0.46} & \num{676.1} & \num{44.24} & \num{0.47} & \num{674} & \num{0.10} & \citet{2010AASegransan} & \citet{2010AASegransan} & \citet{2010AASegransan} \\
HD147018c & \num{1008} & \num{0.13} & \num{2095.6} & \num{1008} & \num{0.13} & \num{2080} & \num{0.10} & \citet{2010AASegransan} & \citet{2010AASegransan} & \citet{2010AASegransan} \\
HD150433b & \num{1096.27} & \ldots & \num{53.48} & \ldots & \ldots & \ldots & \num{-0.36} & \citet{2011MayorArxiv} & \citet{2011MayorArxiv} & \citet{2008AASousa} \\
HD154088b & \num{18.59} & \ldots & \num{5.2} & \num{18.56} & \num{0.19} & \underline{\num{6.6}} & \num{0.28} & \citet{2021AAUnger} & \citet{2021AAUnger} & \citet{2008AASousa} \\
HD156846b & \num{359.51} & \num{0.84} & \num{3498.5} & \num{359.51} & \num{0.85} & \num{3321} & \num{0.22} & \citet{2008AATamuz} & \citet{2008AATamuz} & \citet{2008AATamuz} \\
HD157172b & \num{104.99} & \num{0.33} & \num{37.19} & \ldots & \ldots & \ldots & \num{0.11} & \citet{2011MayorArxiv} & \citet{2011MayorArxiv} & \citet{2008AASousa} \\
HD160691d & \num{9.64} & \num{0.12} & \num{10.6} & \num{9.64} & \underline{\num{0.17}} & \num{10.55} & \num{0.30} & \citet{2004AASantosB} & \citet{2007AAPepe} & \citet{2018AASousa} \\
HD160691e & \num{313.2} & \num{0.04} & \num{189.1} & \num{310.55} & \underline{\num{0.07}} & \underline{\num{165.9}} & \num{0.30} & \citet{2007AAPepe} & \citet{2007AAPepe} & \citet{2018AASousa} \\
HD160691b & \num{648.7} & \num{0.18} & \num{546.9} & \num{643.25} & \underline{\num{0.13}} & \num{532.7} & \num{0.30} & \citet{2001ApJButler} & \citet{2007AAPepe} & \citet{2018AASousa} \\
HD160691c & \num{4210} & \num{0.43} & \num{790} & \num{4210} & \underline{\num{0.10}} & \underline{\num{576.5}} & \num{0.30} & \citet{2004ApJMcCarthy} & \citet{2007AAPepe} & \citet{2018AASousa} \\
HD166724b & \num{5144} & \num{0.73} & \num{1121.8} & \num{5144} & \num{0.74} & \num{1122} & \num{-0.09} & \citet{2013AAMarmier} & \citet{2013AAMarmier} & \citet{2013AATsantaki} \\
HD168443b & \num{58.11} & \num{0.52} & \num{2446.5} & \num{58.11} & \num{0.53} & \num{2517} & \num{0.06} & \citet{1999ApJMarcy} & \citet{2022ApJSRosenthal} & \citet{2018AASousa} \\
HD168443c & \num{1749.83} & \num{0.21} & \num{5525.7} & \num{1749.67} & \num{0.21} & \num{5645} & \num{0.06} & \citet{2001ApJMarcy} & \citet{2022ApJSRosenthal} & \citet{2018AASousa} \\
HD168746b & \num{6.40} & \num{0.10} & \num{77.9} & \num{6.40} & \underline{\num{0.08}} & \num{73} & \num{-0.10} & \citet{2002AAPepe} & \citet{2002AAPepe} & \citet{2008AASousa} \\
HD169830b & \num{225.62} & \num{0.31} & \num{918.4} & \num{225.62} & \num{0.31} & \num{915.4} & \num{0.18} & \citet{2001AANaef} & \citet{2004AAMayor} & \citet{2008AASousa} \\
HD169830c & \num{2102} & \num{0.33} & \num{1291.7} & \num{2102} & \num{0.33} & \num{1284} & \num{0.18} & \citet{2004AAMayor} & \citet{2004AAMayor} & \citet{2008AASousa} \\
HD179949b & \num{3.09} & \num{0.02} & \num{286.7} & \num{3.09} & \num{0.02} & \num{307} & \num{0.21} & \citet{2001ApJTinney} & \citet{2022ApJSRosenthal} & \citet{2008AASousa} \\
HD181433b & \num{9.37} & \num{0.42} & \num{7.32} & \num{9.37} & \num{0.40} & \num{7.5} & \num{0.36} & \citet{2009AABouchy} & \citet{2009AABouchy} & \citet{2013AATsantaki} \\
HD181433c & \num{1019} & \num{0.25} & \num{222} & \num{962} & \underline{\num{0.28}} & \num{203} & \num{0.36} & \citet{2009AABouchy} & \citet{2009AABouchy} & \citet{2013AATsantaki} \\
HD181433d & \num{3201} & \num{0.11} & \num{184} & \underline{\num{2172}} & \underline{\num{0.48}} & \num{171} & \num{0.36} & \citet{2009AABouchy} & \citet{2009AABouchy} & \citet{2013AATsantaki} \\
HD187085b & \num{986} & \num{0.46} & \num{255.4} & \num{1019} & \underline{\num{0.25}} & \num{266} & \num{0.13} & \citet{2006MNRASJones} & \citet{2018AABarbato} & \citet{2013AASantos} \\
HD189567b & \num{14.28} & \ldots & \num{8.6} & \num{14.29} & \num{<0.10} & \num{8.5} & \num{-0.24} & \citet{2021AAUnger} & \citet{2021AAUnger} & \citet{2008AASousa} \\
HD189567c & \num{37.20} & \ldots & \num{8.3} & \num{33.69} & \num{0.16} & \underline{\num{7.0}} & \num{-0.24} & \citet{2021AAUnger} & \citet{2021AAUnger} & \citet{2008AASousa} \\
HD189567d & \num{61.8} & \ldots & \num{10.4} & \ldots & \ldots & \ldots & \num{-0.24} & \citet{2021AAUnger} & \citet{2021AAUnger} & \citet{2008AASousa} \\
HD192310b & \num{74.72} & \num{0.13} & \num{16.9} & \num{74.72} & \num{0.13} & \num{16.9} & \num{-0.01} & \citet{2011ApJHowardA} & \citet{2011AAPepe} & \citet{2017AAAndreasen} \\
HD192310c & \num{525.8} & \num{0.32} & \num{24} & \num{525.8} & \num{0.32} & \num{24} & \num{-0.01} & \citet{2011AAPepe} & \citet{2011AAPepe} & \citet{2017AAAndreasen} \\
HD196050b & \num{1378} & \num{0.22} & \num{1321} & \num{1378} & \underline{\num{0.30}} & \underline{\num{960}} & \num{0.23} & \citet{2002MNRASJones} & \citet{2004AAMayor} & \citet{2008AASousa} \\
HD196067b & \num{3638} & \num{0.66} & \num{1595.4} & \num{3638} & \num{0.66} & \ldots & \num{0.23} & \citet{2013AAMarmier} & \citet{2013AAMarmier} & \citet{2013AASantos} \\
HD204313c & \num{34.91} & \num{0.19} & \num{16.86} & \num{34.91} & \underline{\num{0.16}} & \num{17.6} & \num{0.18} & \citet{2016AADiaz} & \citet{2016AADiaz} & \citet{2008AASousa} \\
HD204313b & \num{2036.20} & \num{0.09} & \num{1398.82} & \num{2024.1} & \num{0.09} & \num{1360} & \num{0.18} & \citet{2010AASegransan} & \citet{2016AADiaz} & \citet{2008AASousa} \\
HD204313e & \num{11949} & \num{0.37} & \num{516.11} & \underline{\num{7325}} & \underline{\num{0.25}} & \underline{\num{4868}} & \num{0.18} & \citet{2022ApJSFeng} & \citet{2022ApJSFeng} & \citet{2008AASousa} \\
HD204941b & \num{1733} & \num{0.37} & \num{33.22} & \num{1733} & \num{0.37} & \underline{\num{85}} & \num{-0.19} & \citet{2011AADumusque} & \citet{2011AADumusque} & \citet{2011AADumusque} \\
HD208487b & \num{130.08} & \num{0.23} & \num{162.8} & \num{130} & \underline{\num{0.32}} & \underline{\num{126}} & \num{0.08} & \citet{2005ApJTinney} & \citet{2005ApJTinney} & \citet{2008AASousa} \\
HD210277b & \num{442.19} & \num{0.47} & \num{404.5} & \num{442.94} & \num{0.47} & \num{392.8} & \num{0.18} & \citet{1999ApJMarcy} & \citet{2022ApJSRosenthal} & \citet{2008AASousa} \\
HD213240b & \num{882.7} & \num{0.42} & \num{1440.6} & \num{951} & \num{0.45} & \num{1430} & \num{0.14} & \citet{2001AASantosB} & \citet{2001AASantosB} & \citet{2008AASousa} \\
HD215152c & \num{7.28} & \num{0.34} & \num{2.77} & \num{7.28} & \num{0.34} & \num{2.77} & \num{-0.08} & \citet{2018AADelisle} & \citet{2018AADelisle} & \citet{2013AATsantaki} \\
HD215152d & \num{10.86} & \num{0.38} & \num{3.09} & \num{10.86} & \num{0.38} & \num{3.09} & \num{-0.08} & \citet{2018AADelisle} & \citet{2018AADelisle} & \citet{2013AATsantaki} \\
HD215456b & \num{192.74} & \num{0.01} & \num{31.98} & \ldots & \ldots & \ldots & \num{-0.09} & \citet{2011MayorArxiv} & \citet{2011MayorArxiv} & \citet{2008AASousa} \\
HD215456c & \num{2263.9} & \ldots & \num{76.04} & \ldots & \ldots & \ldots & \num{-0.09} & \citet{2011MayorArxiv} & \citet{2011MayorArxiv} & \citet{2008AASousa} \\
HD216435b & \num{1311} & \num{0.07} & \num{386.1} & \num{1311} & \num{0.07} & \num{400.4} & \num{0.24} & \citet{2003MNRASJones} & \citet{2006ApJButler} & \citet{2008AASousa} \\
HD216437b & \num{1353} & \num{0.31} & \num{689.1} & \num{1334} & \num{0.32} & \num{706.5} & \num{0.24} & \citet{2002MNRASJones} & \citet{2018AABarbato} & \citet{2018AASousa} \\
HD216770b & \num{118.45} & \num{0.37} & \num{205.6} & \num{118.45} & \num{0.37} & \num{207} & \num{0.24} & \citet{2004AAMayor} & \citet{2004AAMayor} & \citet{2008AASousa} \\
HD217107b & \num{7.12} & \num{0.12} & \num{445.3} & \num{7.12} & \num{0.13} & \num{440.2} & \num{0.34} & \citet{1999PASPFischer} & \citet{2022ApJSRosenthal} & \citet{2018AASousa} \\
HD217107c & \num{4270} & \num{0.51} & \num{831.3} & \underline{\num{5141}} & \underline{\num{0.39}} & \underline{\num{1370}} & \num{0.34} & \citet{2005ApJVogt} & \citet{2022ApJSRosenthal} & \citet{2018AASousa} \\
HD218566b & \num{225.7} & \num{0.30} & \num{67.7} & \num{225.2} & \underline{\num{0.27}} & \num{62.9} & \num{0.17} & \citet{2011ApJMeschiari} & \citet{2022ApJSRosenthal} & \citet{2013AASantos} \\
HD219077b & \num{5501} & \num{0.77} & \num{3301.94} & \num{5501} & \num{0.77} & \num{3302.11} & \num{-0.13} & \citet{2013AAMarmier} & \citet{2013AAMarmier} & \citet{2008AASousa} \\
HD220689b & \num{2209} & \num{0.16} & \num{336.87} & \num{2266} & \underline{\num{0.05}} & \num{355.3} & \num{-0.01} & \citet{2013AAMarmier} & \citet{2018AABarbato} & \citet{2013AASantos} \\
HD222582b & \num{572.38} & \num{0.72} & \num{2425.1} & \num{572.54} & \num{0.76} & \num{2504} & \num{-0.01} & \citet{2000ApJVogt} & \citet{2022ApJSRosenthal} & \citet{2008AASousa} \\
\hline
\end{longtable}
\end{landscape}

\end{appendix}

\end{document}